\documentclass{aa}
\usepackage{epsfig}
\usepackage{graphicx}
\usepackage{psfig,epsf}

\def\cm3{cm$^{-3}$}
\def\kms{km~s$^{-1}$}

\def\rsun{R$_{\odot}$}

\def\beq{\begin{equation}}
\def\eeq{\end{equation}}

\begin{document}

\title{Quantitative Spectroscopy of Photospheric-Phase
Type II Supernovae}

\subtitle{}

\author{Luc Dessart\inst{1,3},
        \and
        D. John Hillier\inst{2}
        }
\offprints{Luc Dessart,\\ \email{luc@as.arizona.edu}}

  \institute{Max-Planck-Institut f\"{u}r Astrophysik,
             Karl-Schwarzschild-Str 1, 85748, Garching bei M\"{u}nchen, Germany 
        \and
           Department of Physics and Astronomy, University of Pittsburgh,
           3941 O'Hara Street, Pittsburgh, PA, 15260 
        \and
           Steward Observatory, University of Arizona, 933 North Cherry Avenue, 
        Tucson, AZ 85721, USA 
            }

\date{Accepted/Received}

\abstract{
We present first results on the quantitative spectroscopic analysis of the photospheric-phase
of Type II supernovae (SN). The analyses are based on the model atmosphere code,
CMFGEN, of Hillier \& Miller (1998)
which solves the radiative transfer and statistical equilibrium equations in expanding
outflows under the constraint of radiative equilibrium.
A key asset of CMFGEN is its thorough treatment of line-blanketing due to metal species.
From its applicability to hot star environments, the main modifications to the source
code were to allow a linear velocity law, a power-law density distribution,
an adaptive grid to handle the steep H recombination/ionization front
occurring in some SN models, and a
routine to compute the gray temperature structure in the presence of large velocities.
In this first paper we demonstrate the ability of CMFGEN to reproduce,
with a high level of accuracy, the UV and optical observations of a sample of well observed
Type II SN, i.e. SN1987A and SN1999em, at representative stages of their photospheric
evolution. Two principal stages of SN are modeled --- that
where hydrogen is fully ionized, and that in which H is only partially ionized.
For models with an effective temperature below $\sim$8,000K, hydrogen recombines and gives
rise to a steep ionization front.

The effect of varying the location of the outer grid radius
on the spectral energy distribution (SED) is investigated.
We find that going to 5-6 times the optically-thick base radius is optimal,
since above that, the model becomes prohibitively large,
while below this, significant differences appear because of reduced line-blanketing
(which persists even far above the photosphere) and the truncation of line-formation
regions. To constrain the metallicity and the reddening of SN, the
UV spectral region of early-time spectra is essential.
We find that the density of the photosphere and effect of line blanketing
decline as the spatial scale of the SN increases.
The density distribution is found to have a strong impact on the overall flux distribution
as well as line profiles.
For a given base density, the faster the density drops, the higher the effective
temperature of the model.
We also find in cool models that the set of Ca{\,\sc ii} lines, near 8500\AA\ is strongly
sensitive to the density gradient. They show a weaker and narrower profile for steeper
density distributions.
Hydrogen Balmer lines are very well reproduced in fully or partially
ionized models, but underestimated when hydrogen recombines.
A reduced turbulent velocity or a flatter density layout are found to partially,
but not fully, cure this persistent problem in studies of Type II SN.
He{\,\sc i} lines observed in early-time spectra are very well reproduced, even for
very modest helium enrichments, likely resulting from treatment of important non-LTE effects.
At similar early epochs CMFGEN predicts, unambiguously, the presence of N{\,\sc ii} lines in the
blue-wing of both H$\beta$ and \ion{He}{\,\sc i} 5875\AA. These lines have been
observed but so far have generally been associated with
peculiar emission, from locations far above the photosphere,
in the strong adjacent lines.

Finally, we present a pedagogical investigation on P-Cygni profile formation in
Type II SN.
H$\alpha$ is found to form very close to the photosphere and thus presents a significant
flux-deficit in the red, made greater by the rapidly declining density distribution.
This provides a clear explanation for the noticeable blue-shift of P-Cygni profiles
observed in early-time spectra of Type II SN.

Future studies based on CMFGEN modeling will focus on using Type II SN for the
calibration of distances in the Universe, as well as on detailed spectroscopic analyses
for the determination of progenitor properties.


}
\titlerunning{Type II supernovae spectroscopic modeling}

\maketitle

\section{Introduction}
\label{Sec_intro}

Following the development of radiative transfer algorithms suitable for expanding
outflows (Mihalas et al. 1975), and advances in computer technology, increasingly detailed
and successful quantitative spectroscopic analyses of massive stars and their
winds have been carried out (c.f., Schmutz 1997; Hillier \& Miller 1998, 1999;
Hamann \& Koesterke 1998).
Such quantitative spectroscopic studies can shed light on numerous properties of
optically-thick outflows, with implications for understanding
the parent object.  One can determine the chemical composition, radius, luminosity, density,
ionization structure and kinematic properties of the star and its wind, as well as the
temperature of the equivalent blackbody at the thermalisation depth.

Although stellar winds are radiatively-driven, the
mechanisms controlling the escape of photons from the optically-thick
flow are similar in nature to that occurring in the freely-expanding fast
ejecta associated with supernovae (SN) explosions, occurring either {\it via }
core-collapse of a massive star (H-free Type Ib/c or H-rich Type II;
Arnett et al. 1989) or thermonuclear runaway
of a Chandrasekhar mass white-dwarf (Type Ia; Hillebrandt \& Niemeyer 2000).
In the context of SN, spectroscopic modeling allows inferences on the properties of
the progenitor of the explosion to be made, providing
clues to the evolution and death of stars, as well as the formation of compact objects.

Additionally, massive stars, and SN explosions to a much larger extent, are
so intrinsically luminous that attempts to use them as distance calibrators have been made.
For example, massive stars can be used to constrain the Hubble constant via the
Wind-Momentum-Luminosity relation (Kudritzki et al. 1999).
Light curves of Type Ia SN have been used to claim that we live in an
accelerating universe (Riess et al. 1998).
Finally, although Type II SN are not as bright as Type Ia events,
they can be used with variations on the Baade's method (Baade 1926), i.e.
the Expanding Photosphere Method (EPM; Kirshner \& Khan 1974, Eastman \& Kirshner 1989,
Eastman et al. 1996, Schmidt et al. 1994ab, Leonard et al. 2002ab, Hamuy \& Pinto 2002)
and the Spectral-fitting Expanding Atmosphere Method (SEAM; Baron et al. 1995, 2004),
to provide distance estimates for objects even within the Hubble flow.
However, such distance determinations will benefit from a thorough modeling of the SN spectrum
and light curve, as well as from an understanding of the physics of spectrum formation.

Model atmospheres and techniques, which have been tested so heavily and applied so
successfully to hot luminous stars, constitute excellent tools
for such SN analyses.
This is being done with a number of codes: WM--BASIC (Pauldrach et al.
2001, Pauldrach et al. 1996) PHOENIX (Hauschildt 1992ab; Hauschildt
\& Baron 1995), EDDINGTON (Eastman \& Pinto 1993).
Our analyses are based on CMFGEN (Hillier \& Miller 1998),
which has been used to model OB stars, Luminous Blue Variables (LBVs) and
Wolf-Rayet (WR) stars (Hillier et al. 1998, Dessart et al. 2000,
Crowther et al. 2002, Hillier et al. 2003).

O star winds are weak and scattering dominated, quite like SN outflows.
In the weakest-wind O-stars, line formation occurs over a small spatial
region so that sphericity effects are unimportant.
On the contrary O-supergiants, where sphericity effects are important,
show a range of line formation processes, with both
photospheric and wind sites of emission, absorption and scattering
(Hillier et al. 2003, Martins et al. 2004, Crowther et al. 2002).
Conversely, LBV winds are very extended, dense and relatively cool,
and have an ionization structure that
resembles quite closely that of SN. Processes such as Rayleigh
scattering, charge exchange and two-photon decay are of relevance
in both LBV and SN outflows
(Hillier et al. 1998, Hillier et al. 2001, Najarro et al. 1997).
WR winds are dense and fast, and typically show a much higher ionization
than SN. Their
composition is typical of progenitors of Type II (H-rich) and
Type Ib/c (H-free) SN (Dessart et al. 2000, Crowther et al. 2002).

Despite the similarities, SN conditions depart from those of hot star winds
in a number of ways.
The flow is accelerated only temporarily, say for about a day after core-collapse
or explosion, followed by a homologous expansion in which the velocity follows
a Hubble law ($v/r$ constant - Woosley 1988).
Consequently the outflow density profile does not follow an inverse square law as
imposed by mass conservation (Castor, Abbott \& Klein 1975), but instead results from the
disruption of the envelope by the shock wave arising from the explosion
(Woosley 1988).
Indeed models of such explosions, as well as spectroscopic analyses, support a very fast
decreasing density. This rapid density decline, together
with the large velocity gradient, are
key ingredients controlling the escape of radiation from the flow.

Another key component for the modeling of outflows is the origin of opacity.
In H-rich environments, such as massive star winds and Type II SN,
the dominant source of continuum opacity is electron scattering.
Although electron scattering is more important in Type II SN,
there is still significant true absorption, due to bound-free and free-free processes,
to allow the diffusion approximation, and
the assumption of LTE, to be applied at the base of the outflow.
For H-free conditions, bound-free and bound-bound transitions of metals will become
the dominant source of opacity.
There, including as many levels and lines is mandatory for recovering LTE and
a thermalisation layer at the base.

Because of their closer similarity to hot massive stars,
we concentrate on Type II SN in this first series of papers.
In this first paper, we present the necessary adjustments made to CMFGEN in order
to model the emergent spectra from SN outflows, and provide some
key insights into SN spectrum formation.
In Sect.~\ref{Sec_pres_code}, we present the code of Hillier \& Miller (1998), giving the
necessary background and highlighting key differences between CMFGEN
and other models developed and used by different groups.
We also describe the adjustments from hot star to SN conditions.
In Sect.~\ref{Sec_spec_mod}, we present typical results of our spectroscopic modeling,
selecting a sequence of observations taken during the photospheric-phase
evolution of SN1987A and SN1999em. These serve as illustrations for the
model behavior, as well as a basis for our interpretation of key features
of Type II SN during the photospheric phase.
We also give model parameters, discuss outflow properties and
provide line identifications.
In Sect. \ref{Sec_mod_res}, based on the models shown in Sect.~\ref{Sec_spec_mod}
as well as the large parameter space investigated for the preparation of this paper,
we describe generic features of
Type II SN and how these are sensitive to model parameters.
We discuss the handling of the H-recombination front in cool
models (Sect.~\ref{Sec_H_recom}), the potential problem with the choice of outer grid location
(Sect.~\ref{Sec_out_bound}), the importance of line-blanketing (Sect.~\ref{Sec_inf_met}),
the influence of the luminosity (Sect. \ref{Sec_inf_lum}) and density exponent
(Sect.~\ref{Sec_den_exp}) on the spectral morphology, and the modeling of hydrogen, helium
and nitrogen line profiles in the optical range (Sect. \ref{Sec_sel_lines}).
In Sect.~\ref{Sec_line_form}, we describe in detail the process of line formation in Type II
SN, giving in particular a physical explanation for the observed blue-shift of Balmer
lines in photospheric-phase spectra of Type II SN.
In Sect.~\ref{Sec_conc}, we present our conclusions and map out our future endeavors.

\section{Presentation of the code}
\label{Sec_pres_code}

Our effort to analyze the light emanating from SN outflows makes use of
the model atmosphere code CMFGEN, discussed in detail by Hillier \&
Miller (1998, 1999).
CMFGEN solves the radiative transfer equation in the comoving frame, subject
to the constraints of radiative and statistical equilibrium.
The outflow is assumed to be steady-state and to have a spherical symmetry.
The general procedure for analyzing a given object is to input global
characteristic quantities such as luminosity, radius, chemical composition, velocity
and density.

In hot star outflows, the velocity law is taken from detailed hydrodynamical
computations of radiatively-driven winds.
For SN outflows, after about one day after the explosion, the outflow
no longer accelerates but instead freely-expands into the surrounding medium.
Such an homologous expansion is characterized by a Hubble velocity law,
$v(r) = v_0\,(r/R_0)$
where $R_0$ is the base radius where the luminosity
is input, and $v_0$ the velocity of this layer.
This layer in Type II SN is chosen such that it is optically thick at all wavelengths
and thus corresponds to a thermalisation layer where photons have
a Planckian distribution at the local temperature
$T= (L_{\ast}/4 \pi \sigma R_0^2)^{1/4}$, $\sigma$ being Stefan's constant.
Note that this lies significantly below the photosphere which we define
as the layer where the total continuum optical-depth is 2/3 (this includes
bound-free and free-free processes from all species plus electron-scattering), located
typically around 1.3--1.5 $R_0$.
As time proceeds, (constant-velocity) mass shells expand and become more
optically thin, so that in the frame comoving with a given mass shell, the
thermalisation layer recedes deeper into the flow.
At present, the time evolution is taken into account by fitting not only the
spectral energy distribution (SED) but also the absolute flux level at all times.
The former gives a tight constraint on the ionization balance of the flow and
the temperature of the thermalisation layer.
The second constrains the luminosity (or radius of the thermalisation layer).
In practice, we vary the outflow ionization and the absolute flux level
by adjusting the radius scale of the model, controlled by the parameter $R_0$,
and the luminosity (or the flux) going through this inner model boundary,
controlled by the parameter $L_{\ast}$.

For the distance, which is required for the luminosity constraint,
we have adopted the Cepheid or EPM distance to the object under investigation.
As discussed later in this paper, for a given outflow ionization,
the SED is only sensitive to large changes
in luminosity so uncertainties in the distance, at the 50\% level, will
have negligible effects on the SED.

We adopt an analytical density law and adjust it at different epochs
according to whatever nature demands.
Later we show that the steepness of the density fall-off is
mostly constrained by P-Cygni line profile shapes.
The analytical form adopted is a power law, i.e. $\rho(r) = \rho_0 (R_0/r)^n$.
Former investigations, both of the explosion itself as well as others
on the SED have shown that $n$ is large, with a nominal value of 10,
and tends to decrease over time (see, e.g., Arnett 1988, Lucy 1987,
Eastman \& Kirshner 1989).
$\rho_0$ is adjusted so that the Rosseland-mean or electron-scattering
optical depth at the base is of the order of 50.
This ensures photons are thermalised in the base layer.

In radiatively-driven winds of hot stars, the density is
usually taken to be clumped on small scales since this is expected both
theoretically (Owocki \& Rybicki 1984, Dessart \& Owocki 2003)
and observationally (Robert 1992).
In multi-dimensional simulations of SN explosions
(Kifonidis et al. 2003, Reinecke et al. 2002), density contrasts
in the heterogeneous outflows are of the order of a few at most,
thus more than an order of magnitude less than suspected in hot stars.
Hence, we assume no clumping of the outflow in our simulations.
As we show in the next section, the quality of the fits is excellent,
suggesting that clumping, if present, remains small.

In addition to clumping, SN outflows are expected to exhibit turbulence
on a variety of scales (see, e.g., Kifonidis et al. 2003, Burrows et al. 1995).
We generally adopt a microturbulent velocity of 100\,km\,s$^{-1}$, which 
is just a few percents 
of the photospheric velocity -- this has the effect of broadening the
intrinsic (Gaussian) line profile width, increasing the probability
of line-overlap but reducing the required frequency sampling for the
radiative-transfer solution.
Large scale turbulent motions are ignored.

In Type II SN, only a small amount of metals is nucleo-synthesized
during the explosion (Woosley \& Weaver 1995), so that it
seems suitable to assume strict radiative-equilibrium.
To ensure that such a situation prevails, we limit our investigation to early
times during the photospheric phase, say the first month after maximum B band
light.

A key asset of CMFGEN is that all lines of all species are treated in non-LTE.
Since adequate modeling of the blanketing effects of iron are essential,
and since such a metal has a very complex atomic structures with lots of
lines, the atomic structure input in the code is re-organized in terms of
``super-levels" (SL). In this scheme nearby levels, having similar departure coefficients,
are grouped together and treated as an individual level. Despite this grouping,
all lines are treated at their correct wavelengths (ignoring
possible difficulties with the atomic data).
Applying this procedure to all species permits the treatment of many
more ions and bound-bound transitions. In our approach we do not need
to assign an absorption/scattering cross section to the line blanketing.
This is automatically determined by the coupled solution of the
transfer and statistical equilibrium processes. If the SL assignment is
too coarse, branching (i.e., the decay of the upper level to alternate
states) may not be treated correctly. In particular, to treat the Bowen fluorescence
mechanism, special care must be taken when making the SL assignments. However, numerous
tests, in a wide variety of conditions, have shown that the SL approach can yield
an accurate temperature and ionization structure of the envelope. Importantly,
it is relatively easy to change the SL assignments to test their effect.

We show in Table 1 the largest atomic model used in our computations,
showing the number of ``full-" and ``super-" levels, and the corresponding
number of transitions.
For coarse analyses, only H, He, C, N, O and Fe are included.
In fact, including Fe{\,\sc ii} ensures that the overall temperature structure and
ionizing flux are not significantly affected when performing the more
detailed computation with the larger model atom.
The atomic data come from a wide variety of sources,
the Opacity Project (Seaton 1987; The Opacity Project Team 1995, 1997),
the Iron Project (Pradhan et al. 1996; Hummer et al. 1993),
Kurucz (1995), and the Atomic Spectra Database at NIST Physical
Laboratory being the principal sources.
Much of the Kurucz data was obtained directly from the Center for 
Astrophsyics (Kurucz 1988, 2002).
Individual sources of atomic data include the following:
Bautista \&  Pradhan (1997), Becker \& Butler (1995), Butler et al. (1993),
Fuhr, Martin, \& Wiese (1988), Kingdon \& Ferland (1996),
Luo \& Pradhan (1989), Luo et al. (1989),
Mendoza (1983), Mendoza et al. (1995), Nahar (1995, 1996), Nahar \&  Pradhan (1996),
Neufeld \&  Dalgarno (1987), Nussbaumer \& Storey (1983, 1984),
Peach, Saraph, \& Seaton (1988), Storey  (1988), Tully, Seaton \& Berrington (1990),
Wiese, Smith, \& Glennon (1966), Wiese, Smith, \& Miles (1969),
Zhang \& Pradhan (1995, 1997).
Note that at present, we are missing some species for the modeling of late-stage spectra,
most importantly C{\,\sc i}, Sc{\,\sc ii} and Ba{\,\sc ii}, something we will remedy
in the future.

\begin{table}
\caption[]{Summary of the model atom used in our calculations.
$N_{\hbox{\sc f}}$ is the number of full levels, $N_{\hbox{\sc s}}$ the number
of super levels and $N_{\hbox{\sc t}}$ the corresponding number of
transitions.
}
\label{tabatom}
\begin{tabular}{
l@{\hspace{0mm}}
c@{\hspace{0.5mm}}c@{\hspace{0.5mm}}cc@{\hspace{0.5mm}}c@{\hspace{0.5mm}}cc@{\hspace{0.5mm}}c@{\hspace{0.5mm}}c
c@{\hspace{0.5mm}}c@{\hspace{0.5mm}}c}
\hline
\hline
& \multicolumn{3}{c}{I   } &
  \multicolumn{3}{c}{II  } &
  \multicolumn{3}{c}{III } &
  \multicolumn{3}{c}{IV  } \\

Ion          &
$N_{\hbox{\sc f}}$  &  $N_{\hbox{\sc s}}$ & $N_{\hbox{\sc t}}$ &
$N_{\hbox{\sc f}}$  &  $N_{\hbox{\sc s}}$ & $N_{\hbox{\sc t}}$ &
$N_{\hbox{\sc f}}$  &  $N_{\hbox{\sc s}}$ & $N_{\hbox{\sc t}}$ &
$N_{\hbox{\sc f}}$  &  $N_{\hbox{\sc s}}$ & $N_{\hbox{\sc t}}$ \\
\hline
 H  & 30 & 20 &     \\
 He & 51 & 40 & 255& 5  &  5 & 10  \\
  C &    &    &    & 59 & 32 &334  & 20 & 12 & 49 & 14 & 9 & 46 \\
  N & 104& 44 & 760& 41 & 23 &144  &  8 &  8 & 11    \\
  O &  75& 23 & 535&111 & 30 &1093 & 46 & 26 & 184    \\
 Ne &    &    &    &242 & 42 &5397 \\
 Na &  71& 22 & 788& 35 & 21 & 181 \\
 Mg &    &    &    & 65 & 22 &1229 \\
 Al &    &    &    & 44 & 26 & 156 & 45 & 17 &337    \\
 Si &    &    &    & 59 & 31 & 318 & 51 & 27 &235    \\
  S &    &    &    &324 & 56 &7830 & 98 & 48 &832    \\
 Ca &    &    &    & 77 & 21 & 1464  \\
 Ti &    &    &    &152 & 37 &3068 & 206 & 33 &4584  \\
 Cr &    &    &    &196 & 28 &3580 & 145 & 30 &2359  \\
 Mn &    &    &    & 97 & 25 & 210 & 175 & 30 &3173  \\
 Fe &    &    &    &827 &265 &43077& 477 & 61 &6411 & 282 & 50 & 7872 \\
 Co &    &    &    &144 & 34 &2047 & 283 & 41 &6910   \\
 Ni &    &    &    &93  & 19 & 817 & 67  & 15 & 379   \\
\hline
\end{tabular}
\end{table}

Before describing results obtained with CMFGEN, it is worthwhile stressing the
differences with other models used for SN studies.
As described in the introduction, a number of general purpose model atmosphere
codes, often developed originally for the treatment of the expanding outflows
of hot stars, have been applied to SN.
The general practice for the study of Type II SN is to assume
spherical symmetry and steady-state, although some attempts are now made to
perform 3D Monte-Carlo radiative-transfer calculations (Thomas et al. 2003).
The level of refinement for the radiative transfer computation varies
significantly: many studies treat all metal lines in LTE, using non-LTE for
H and He only (Schmutz et al. 1990; Eastman et al. 1989, 1996).
Although much more computationally expensive, a higher level of consistency is
met by studies that treat instead all lines in non-LTE (H\"{o}flich 1988;
Mitchell et al. 2002).
Finally, the alternative parameterized approach with SYNOW, tagged ``direct-analysis'',
is less physically consistent than previous methods, but constitutes a useful
tool to identify spectral features -- a challenging task in SN spectroscopy -- and can
thus serve as a basis for more accurate model atmosphere computations.
%

Before proceeding further it is worth giving an honest appraisal
of some of the deficiencies of the present study. First we assume
spherical geometry and homogeneity. Many spectropolarimetric studies have shown that
at least for some SN, a departure from spherical symmetry on a large scale is present,
although mostly restricted to late times (see, e.g., Kasen et al. 2003 for Type Ia SN,
H\"{o}flich et al. 1996 for Type Ib SN, and Leonard et al. 2001 for Type II SN).
Further, due to Rayleigh-Taylor instabilities,
there is likely to be substantial mixing of material from the inner
layers into the outer layers (Kifonidis et al. 2003), as well as smaller scale
structures (with a scale of the order of or below the photospheric radius value).
One such evidence for inhomogeneities was seen in H$\alpha$ line profile variations
in late-time spectra of SN 1987A (Spyromilio et al. 2003).
Second, we neglect the time dependence of the flow.
This is certainly not valid for modeling the luminosity
evolution from hydrodynamical calculations, especially at early times, but is expected to be
a reasonable approximation in the atmosphere where we can fix the luminosity
using observations.

At present we ignore relativistic effects other than the frequency
shift arising from the first order Doppler correction. Such effects
have been discussed by  Hauschildt et al. (1991), and will be the subject of another paper.
All these effects are crucial for hydrodynamical modeling, but less so in
the approach we adopt where we constrain the emitted luminosity
by the observations, and only treat the ``photospheric'' layers.
Other effects, such as an accurate treatment of line blanketing, are
much more important in determining the observed spectrum.
The primary motivation for doing so was to utilize CMFGEN in a standard configuration
--- this allowed convergence difficulties to be attributable to the
properties of SN envelopes, rather than complications arising from
the relativistic transfer. In an earlier attempt at modeling SN the
strongly scattering envelope gave rise to difficulties in the
temperature convergence which were wrongly attributed to problems in the relativistic
transfer, whereas they were related to the use of SL (Hillier 2003).

\begin{figure*}[htp!]
\epsfig{file=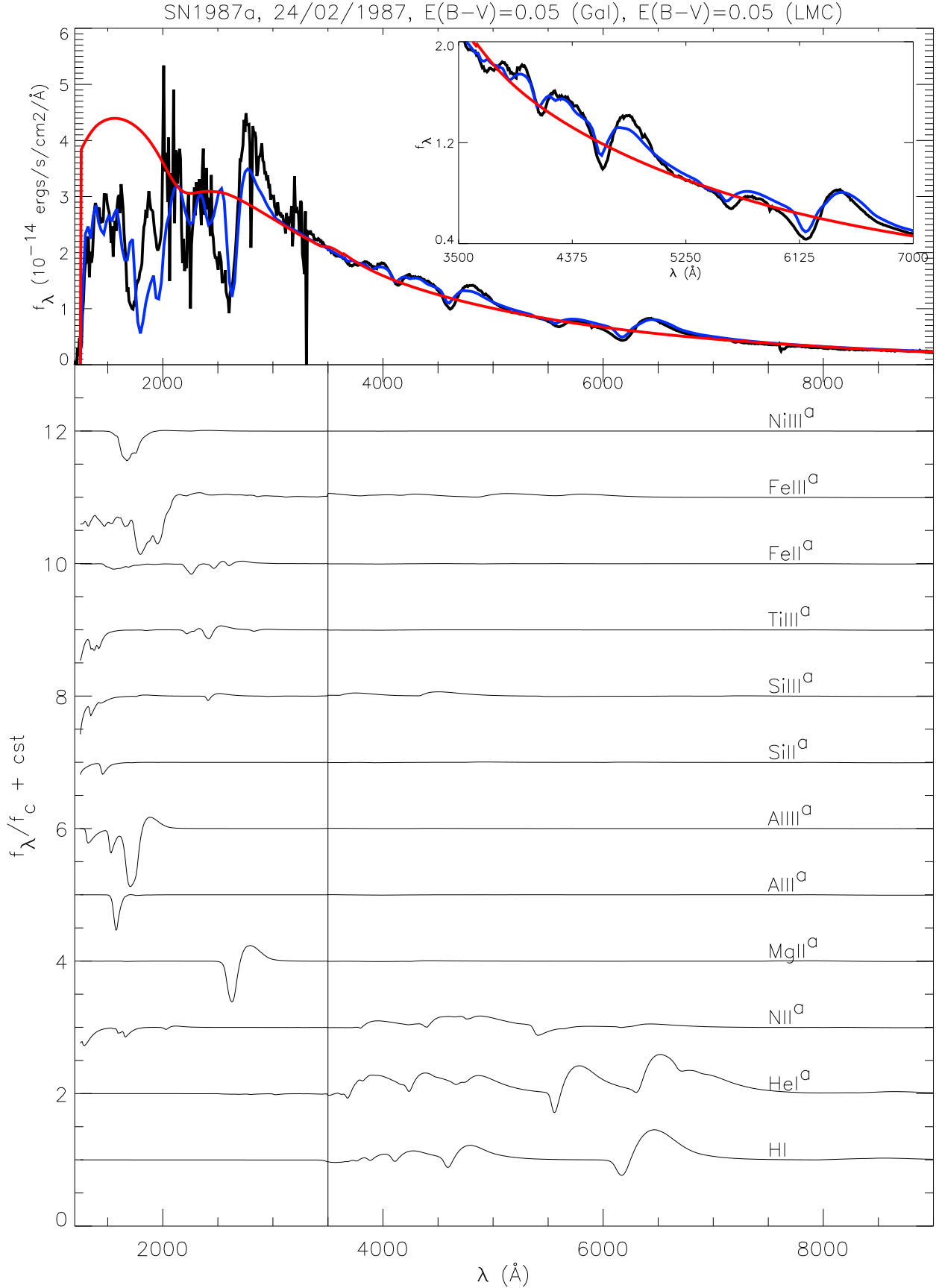, width=18cm}
\caption{
{\bf Top:}  Illustration showing the comparison between reddened model spectra (blue)
and observations of SN1987A (black) taken on the 24th of February 1987
(Pun et al. 1995, Phillips et al. 1988).
Each model flux distribution is scaled for an assumed LMC distance of 52kpc.
In red we show the theoretical continuum energy distribution of the
model.
Model characteristics are: $L_{\ast} = 3 \times 10^8 L_{\odot}$,
$R_{\rm phot} = 2.74 \times 10^{14}$ cm (or 3940 \rsun),
$v_{\rm phot}=17,700$\,km\,s$^{-1}$, $T_{\rm eff} =11,200$K, $n=12$,
$\rho_{\rm phot} = 1.1 \times 10^{-13}$ g\,cm$^{-3}$.
{\bf Bottom:}
Montage of rectified spectra computed by including
bound-bound transitions of individual species,
ordered from the bottom by increasing atomic weight.
For species labeled with an ``a'', we show $(f_{\lambda}/f_{\rm
c}-1) \times xscale+1$, with $xscale=4$ beyond 3500\AA\, and unity elsewhere [color].
}
\label{fig_sn87A}
\end{figure*}

Due to the large photospheric velocities, we modified the gray
temperature formulation to take into account the first order
Doppler shifts. This formulation was necessary to provide an
improved estimate of the temperature structure at depth
at the beginning of the model iteration.

\section{Spectroscopic modeling}
\label{Sec_spec_mod}

In the present paper, we wish to discuss the basic properties of a few models
whose synthetic spectrum fits reasonably well the observations at early, intermediate
and late times of the photospheric-phase evolution of Type II SN.
We base our discussion on two very well observed Type II SN, SN1987A and SN1999em.
The former was associated with the neutrino detection on February 23.316 UT 1987 with
the Kamiokande detector (Koshiba 1987).
Here, we use the first spectrum taken both in the UV and in the optical
(Pun et al. 1995, Phillips et al. 1988), on February 24th, thus only one day
after discovery (and collapse). We use this observation for the early-stage description of
Type II SN spectra.
For SN1999em, no neutrino detection was made and thus no precise dating of
the core-collapse of the progenitor exists.
Inferences, based on the light curve appearance or the EPM, suggest
the explosion took place at around October 27th 1999, with an uncertainty of 1-2 days.
It was absent in images of the host galaxy NGC\,1637 taken on October 20.45 UT,
while first detection of the SN was made on October 29th (JD 2451480.94; Li 1999).
The resemblance of the spectrum taken on October 30th with the spectrum of SN1987A
taken on the 24th of February 1987, indicate that SN1999em was indeed caught at a very early
phase of evolution.
Here, we use observations of SN1999em taken on the 5th of November 1999 (ca. one
week after explosion; Baron et al. 2000, Leonard et al. 2002a) and on the 14th of
November 1999 (ca. two weeks after explosion; Hamuy et al. 2001)
to describe respectively the typical intermediate- and late-stage appearance
of Type II SN spectra.
Such time qualifications to observations are to be taken in a broad context.
We defer until a follow-up paper a detailed analysis with a fine time-sampling
of the evolution of the SED of a Type II SN during the entire photospheric phase.

Let us describe the spectral evolution in the UV/optical ranges of
these two well observed SN.
For each SN, the first spectrum taken shares striking similarities: the flux in the UV is
much larger than in the optical where one sees hydrogen Balmer lines and He{\,\sc i}\,5875\AA.
Following this stage, as the outflow expands and cools, the UV flux diminishes and
the He{\,\sc i}\,5875\AA\,P-Cygni profile weakens.
Depending on the object, this takes a range of time scales:
SN1987A did that step in just one day, while the corresponding spectral evolution
took 10 days for SN1999em.
One can then identify a cooler stage where both the UV flux and He{\,\sc i} lines are completely gone.
Metal lines start appearing in the optical range and the Ca{\,\sc ii} multiplet
around 8500\AA\,is present.
Further, the Na{\,\sc i} doublet at 5890\AA\, starts appearing as a weak P-Cygni profile.
During this entire sequence of events, hydrogen recombination becomes more and more
pronounced so that the next stage is characterized by the presence of the hydrogen
recombination front close to the photosphere.
The SED (i.e., f$_\lambda$) then peaks at or beyond 4000\AA.
Subsequent evolution is more gradual, with significant changes taking place
over longer time-scales.

When fitting synthetic spectra to observations, we redden the theoretical SED
using the Cardelli law (Cardelli et al. 1988) for SN1999em and the Seaton (1979) and
Howarth (1983) laws for the Galaxy and the LMC respectively.
For SN1987A, we take the Cepheid distance to the LMC of 52kpc for SN1987A
and adopt a reddening value $E(B-V)=0.05$ for both galactic and LMC components.
This is slightly less than the value 0.16 proposed by Lundqvist \& Fransson (1996) but seems
warranted to reproduce the observations of SN1987A on the 24th of February.
For later times, this choice has essentially no effect on the fit quality to
observations.
For SN1999em, we use the Cepheid-based distance of 11.7Mpc to
NGC\,1637, somewhat different from the EPM based distance of 8.2Mpc (Leonard et al. 2002a)
or 7.5Mpc (Hamuy et al. 2001).
These differ slightly but within such limits, the computed
SED remains identical, only the absolute flux level to reproduce observations
will change, by about a factor of two.
Note finally that we correct for the recession velocity of the galaxy host of SN1999em
by blue-shifting the observations by 770\,km\,s$^{-1}$ (Leonard et al. 2002a,
Baron et al. 2000).

Due to the ubiquitous presence of line-overlap, it is instructive to
investigate which lines contribute to observed ``line features".
Except on very rare occasions and/or for very few lines
(early-spectra in the optical,
H$\alpha$ for most epochs), the very large expansion velocities are such
that observed features are composed of different lines of different species.
If these arise from species with differing ionization potential, they are
likely to form in distinct regions of the outflow and therefore have a profile
width and strength quite unpredictable a-priori.
To assess the importance of this issue, our code can compute the emergent
spectrum from a formal solution of the transfer equation based on the full
non-LTE solution, including all continuum processes (all bound-free and free-free
processes) but including only the line transitions of the desired species/ionization stage.
Although this then ignores the non-linear interaction of overlapping
lines on the observed spectrum, it allows us to assess with enough accuracy
what species contribute to a given feature.

Finally, in each of the following sections, we give details on model
parameters corresponding to the synthetic spectra shown in the figures.
However, unless otherwise stated, all simulations use a chemical composition
adequate for a young supergiant massive star.
In practice, we use the following CNO-cycle equilibrium values
for a massive progenitor star (Prantzos et al. 1986): slightly enriched
abundances of helium and nitrogen H/He = 5, N/He = 6.8 $\times$ 10$^{-3}$, and
depletion of C and O, i.e. C/He $=$ 1.7 $\times$ 10$^{-4}$ and O/He $=$ 10$^{-4}$
(all given by number).
Additionally, when the model refers to SN1999em, the abundance of metal species
(excluding CNO elements) is set to its Solar Neighborhood.
This is compatible with the inferred solar or over-solar metallicity of
its host galaxy NGC\,1637 (Leonard et al. 2002a).
For SN1987A, metal abundances are all scaled by a fixed factor of 0.4, compared
to their solar value.

\subsection{Early stage}
\label{Sec_early_stage}

\begin{figure}[htp!]
\epsfig{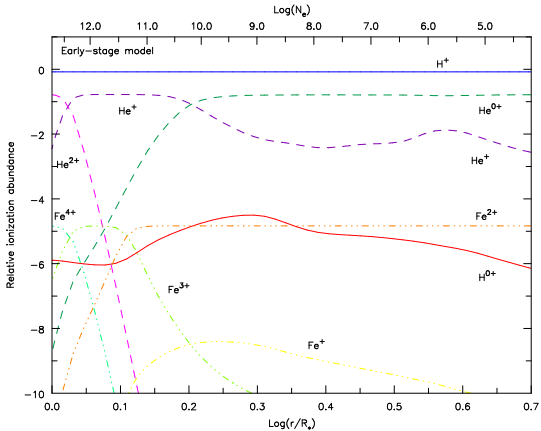}
\caption{
Relative ionization for hydrogen (solid), helium (dashed) and iron (dash-dotted line)
for the early-stage model of Sect.~\ref{Sec_early_stage} and Fig.~\ref{fig_sn87A}.
The ionization fractions are expressed relative to the abundance of all species.
Also shown is the electron density on the top axis [color].
}
\label{fig_sn87A_ion}
\end{figure}

We show in the top panel of Fig.~\ref{fig_sn87A} synthetic fit (color) to observations (black)
of SN1987A taken on the 24th of February 1987.
The red curve represents the continuum energy distribution of the model, while the blue curve
corresponds to the synthetic flux distribution when all line and continuum
processes are included.
The model characteristics are the following:
we have $L_{\ast} = 3 \times 10^8 L_{\odot}$,
$R_{\rm phot} = 2.74 \times 10^{14}$ cm (or 3940$R_{\odot}$),
$v_{\rm phot}=17,700$\,km\,s$^{-1}$, $T_{\rm eff} =11,200$K, $n=12$,
$\rho_{\rm phot} = 1.1 \times 10^{-13}$ g\,cm$^{-3}$.
This model is computed using the full model atom described in Table 1.
We had trouble fitting both UV and optical spectral regions, with the
additional constraint posed by line profiles.
Here, (enhanced) UV flux was partly achieved by adopting a slightly
lower reddening than usually employed (0.1 compared to 0.16).
Obtaining weak absorption and emission in P-Cygni profiles
could only be done by choosing a high density exponent, say above 8,
while much higher values (of the order of 20) led to very weak and anomalous
P-Cygni absorption.
Density exponent values of 10-12 seem best suited for
optical fits to line profiles in this spectrum,
somewhat higher than the value of 6-7 proposed by Lucy (1987).
Admittedly, our fit to the H$\alpha$ trough and H$\beta$ peak flux
could be improved (see inserted box in top-right corner in
Fig.~\ref{fig_sn87A}); a more gradual density fall-off in the regions
where these lines form would likely strengthen the hydrogen Balmer lines
and extend the associated P-Cygni troughs (see Sect.~\ref{Sec_line_form}).

The outflow ionization is high with a SED that peaks in the UV.
The line-blanketing due to metal species is modest --- a significant
UV flux is still present.
We show in the bottom panel of Fig.~\ref{fig_sn87A} the actual contribution
from each of the species present, by computing the synthetic spectrum
including only the bound-bound transitions of a given species.
The resulting SED is then normalized by the continuum energy distribution
and ordered in Fig.~\ref{fig_sn87A} by increasing atomic weight, starting from the bottom.
To enhance the visibility of weak optical features, that nonetheless contribute
to observed features, we apply a scaling to the rectified spectra so that we
in fact show $(f_{\lambda}/f_{\rm c}-1) \times xscale+1$, with $xscale=4$ beyond
3500\AA, and unity elsewhere.

\begin{figure*}[htp!]
\epsfig{file=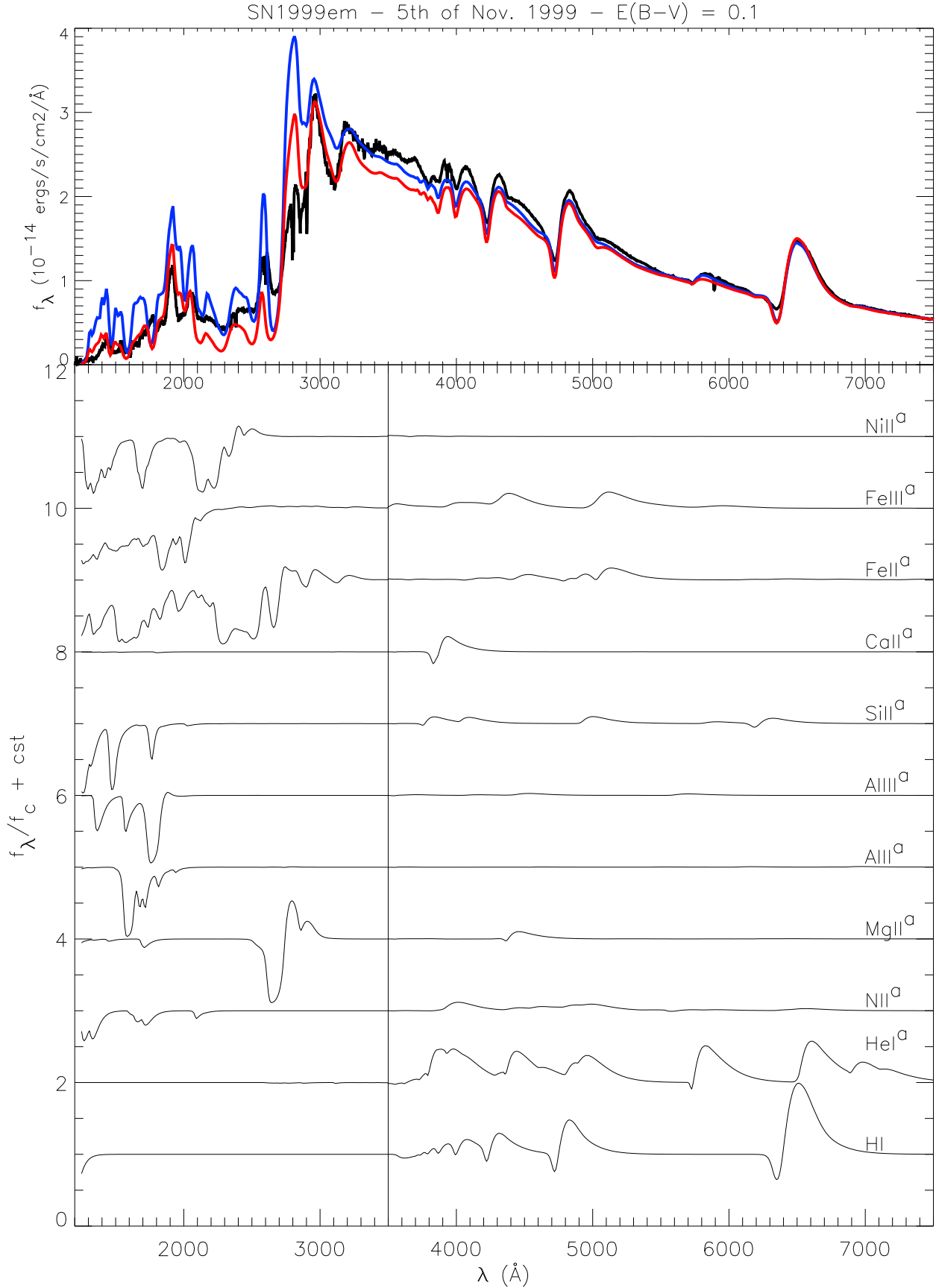, width=18cm}
\caption{
{\bf Top panel:} Synthetic fit (color) to HST and Lick observations (black)
of SN1999em taken on the 5th of November 1999 (Baron et al. 2000, Leonard et
al. 2002a).
The red (blue) curve corresponds to a model with $L_{\ast} = 4 \times 10^8 L_{\odot}$
($L_{\ast} = 5 \times 10^8 L_{\odot}$) and $T_{\rm eff} = 8,850$K ($T_{\rm eff} = 9,200$K).
Other properties are identical:
$R_{\rm phot} = 6.6 \times 10^{14}$ cm (or 9540$R_{\odot}$),
$v_{\rm phot}=8,750$\,km\,s$^{-1}$, $n=10$ and
$\rho_{\rm phot} = 4.1 \times 10^{-14}$ g\,cm$^{-3}$.
{\bf Lower panel:} Montage of rectified spectra (same as in Fig.~\ref{fig_sn87A}) [color].
}
\label{fig_sn99em_5nov}
\end{figure*}

Few features are actually associated with a small set of lines.
At 2800\AA, the feature is due to Mg{\,\sc ii}(3p--2s)\,2802.7\AA\, and the doublet
Mg{\,\sc ii}(4s--3p)\,2928.6--2936.5\AA.
At 1900\AA, the feature is due to a small but larger number of Al{\,\sc iii} lines,
predominantly Al{\,\sc iii}(3p-3s) 1859\AA\, and (4f-3d) 1935\AA.
Other UV features are the result of a large number of lines, in particular
from Fe{\,\sc iii} and Ni{\,\sc iii}.
Indeed, the combination of a rich line-spectrum of metals,
particularly that of iron, together with the large Doppler shifts induce a
strong line-blocking effect on the radiation field.


In the optical (see the inserted zoom in the top right corner of Fig.~\ref{fig_sn87A}),
we find the usual suspects of Type II SN, i.e.
hydrogen Balmer lines: H$\alpha$ at 6562.8\AA, H$\beta$ at 4861.3\AA,
H$\gamma$ at 4340.5\AA, H$\delta$ at 4101.7\AA, and H$\epsilon$ at 3970.0 \AA.
At shorter wavelengths, the higher transitions in the Balmer series merge
together until the Balmer jump at 3646\AA.
The later region might be weakly contaminated by the presence of the Ca{\,\sc ii} doublet
(4p--4s) at 3933.6--3968.5\AA.

We can also find the presence of the He{\,\sc i}\,5875\AA. For SN1987A, this line is only present
in the first spectrum, shown here, while for SN1999em, this line is observed for about a
week (see Sect.~\ref{Sec_hei}).
It serves as a key diagnostic of the outflow ionization since for just slightly
cooler temperatures, it becomes optically-thin and its strength vanishingly
small.
He{\,\sc i}\,5875\AA\, is the one helium line that gets most attention because it is isolated
and thus, uncontaminated. However, the model predicts the presence of a group of He{\,\sc i}
lines around 4000\AA\, (He{\,\sc i}(3p--2s)\,3888.6\AA, He{\,\sc i}(4p--2s)\,3964.7\AA,
He{\,\sc i}(7d--2p)\,4009.2\AA, He{\,\sc i}(7s--2p)\,4024.0\AA, He{\,\sc i}(5d--2p)\,4026.2\AA\,)
as well as around 4500\AA\, (He{\,\sc i}(5d--2p)\,4388.1\AA,
He{\,\sc i}(5s--2p)\,4437.5\AA, He{\,\sc i}(4d--2p)\,4471.5\AA, He{\,\sc i}(4s--2p)\,4713.1\AA),
followed by He{\,\sc i}(4d--2p)\,4921.9\AA\, and He{\,\sc i}(3d--2p)\,6678\AA.
The strength of the resulting features is predicted to be of the same order as
that of He{\,\sc i}\,5875\AA\, and hence, it is clear that helium lines are present
throughout the optical range.


We also find the presence of N{\,\sc ii} lines in the optical:
in the blue wing of He{\,\sc i}\,5875\AA\, as well as in the blue wing of
H$\beta$, at respective wavelengths of 5495--5680\AA\ and 4477--4630\AA\
(each of these is part of a multiplet).
In SN1987A, these features are quite weak (resulting to a large extent from
its unusually high expansion velocity) but their unambiguous presence
is seen in the early spectra of SN1999em (see Sect.~\ref{Sec_nit}) and even more so
in SN1999gi (Leonard et al. 2002b).
We defer until Sect.~\ref{Sec_nit} for a more detailed discussion on these N{\,\sc ii} lines
and their presence in early spectra of Type II SN.
For carbon and oxygen, their expected under-abundance during the
CNO cycle due to mass loss (Prantzos et al. 1986), which we adopt here, is such that not a single
line from C{\,\sc ii}, C{\,\sc iii}, O{\,\sc i}, O{\,\sc ii}, O{\,\sc iii}
is predicted to be noticeable in this model.
Note also that one can see few weak features arising from Si{\,\sc iii} and Fe{\,\sc iii} lines
in the optical, while most metal features are in the UV at this epoch.

Finally, we display the outflow ionization in Fig.~\ref{fig_sn87A_ion}.
Hydrogen is fully ionized, helium doubly ionized only at the base,
singly ionized at intermediate heights and neutral in the outer part of the outflow.
Iron is present in the form of Fe$^{4+}$ inside through to Fe$^{2+}$ outside,
with Fe$^{+}$ never setting in.
Note that in this model, the photosphere is located at $r=1.48 R_0$ or
$\log(r/R_0) = 0.17$, thus in a region where H$^+$, He$^+$ and Fe$^{2+}$ dominate.

\subsection{Intermediate stage}
\label{Sec_int_stage}

In this and the following section, we turn to the subsequent evolution
of Type II SN spectra, using observations performed at ca. 1 and 2 weeks after
the estimated explosion time of SN1999em (Li 1999; see introduction to
Sect.~\ref{Sec_spec_mod}).

\begin{figure}[htp!]
\epsfig{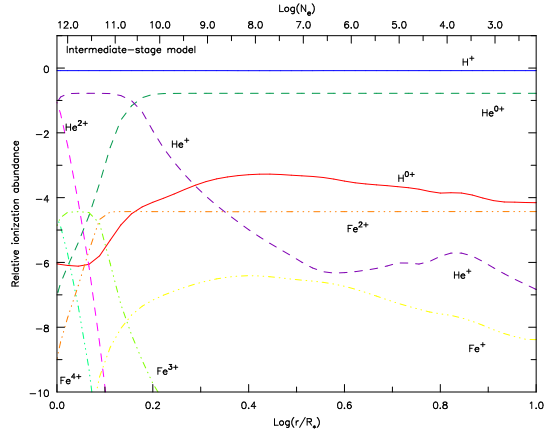}
\caption{
Ionization structure for the intermediate-stage model described in Sect.~\ref{Sec_int_stage}
and shown in Fig.~\ref{fig_sn99em_5nov}. The ionization fractions are expressed
relative to the abundance of all species [color].
}
\label{fig_sn99em_5nov_ion}
\end{figure}

We show in Fig.~\ref{fig_sn99em_5nov} synthetic fits (color) to observations (black)
taken on the 5th of November by Baron et
al. (2000) using the Hubble Space Telescope, as well as optical observations
from Leonard et al. (2002a).
The two models differ only in luminosity: $L_{\ast} = 4 \times 10^8 L_{\odot}$
for the red model and $L_{\ast} = 5 \times 10^8 L_{\odot}$ for the blue model.
The resulting model properties differ only in effective temperature:
$T_{\rm eff} = 8,850$K (red model) and  $T_{\rm eff} = 9,200$K (blue model)
while the common properties (parameters) are the following:
$R_{\rm phot} = 6.6 \times 10^{14}$ cm (or 9540$R_{\odot}$),
$v_{\rm phot}=8,750$\,km\,s$^{-1}$, $n=10$ and
$\rho_{\rm phot} = 4.1 \times 10^{-14}$ g\,cm$^{-3}$.
The red model would give a better fit to observations if we chose
$E(B-V)= 0.08-0.09$ instead of the adopted value of 0.1.
This shows that the UV range is important to place very strong constraints
on the model temperature as well as the adopted reddening, helping to
lift the degeneracy in the impact of these two parameters.

In the bottom panel of Fig.~\ref{fig_sn99em_5nov}, we show the synthetic spectrum for the blue model
when only the bound-bound transitions of selected species (indicated on the right)
are included together with all bound-free and free-free processes.
As in Fig.~\ref{fig_sn87A}, we apply a scaling of a factor of four to the rectified spectra beyond
3500\AA\, to enhance the visibility of weak features.

At this stage of its evolution, the SN outflow has cooled
significantly, so that Fe{\,\sc ii} line-blanketing starts to strongly diminish
the UV flux.
This blanketing is not so severe, so that some residual UV flux is observable, carrying
information on the species at its origin.
From the lower panel, we see that the line-blanketing makes the UV flux
smaller than the continuum flux when all lines are ignored.
In other words, emission features in the UV are to be understood as
spectral windows where this line-blanketing is less relative to
nearby regions, rather than an excess flux above the continuum level.
We thus identify enhanced line-blocking due to Fe{\,\sc ii}, Fe{\,\sc iii}, Ni{\,\sc ii},
Si{\,\sc ii}, Al{\,\sc ii}, Al{\,\sc iii}.
An exception to this is due to Mg{\,\sc ii} lines, already
discussed in the previous section, which show pronounced emission {\it and}
absorption around 2800\AA.

\begin{figure*}[htp!]
\epsfig{file=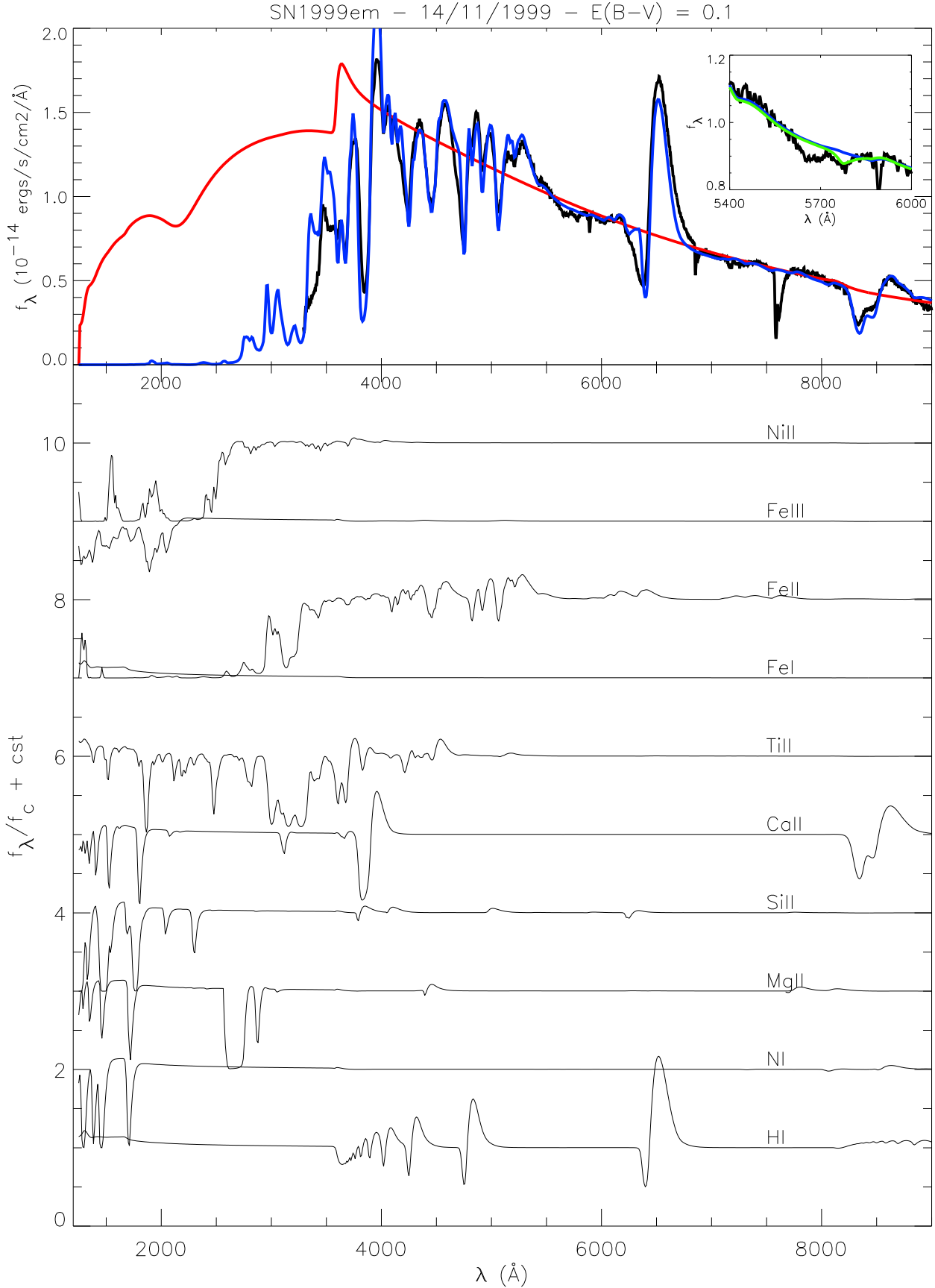, width=18cm}
\caption{
{\bf Top panel:} Synthetic fit (blue) to observations (black) of SN1999em taken
on the 14th of November 1999 (Leonard et al. 2002a). We also show the continuum
flux level (red) computed by ignoring all bound-bound transitions in the formal
solution of the radiative transfer equation.
The model parameters are:
$L_{\ast} = 1.5 \times 10^8 L_{\odot}$,
$R_{\rm phot} = 6.15 \times 10^{14}$ cm (or 8840$R_{\odot}$),
$v_{\rm phot}=6,350$\,km\,s$^{-1}$, $T_{\rm eff} =6,800$K, $n=10$,
$\rho_{\rm phot} = 8.7 \times 10^{-14}$ g\,cm$^{-3}$.
In the top panel, we insert a zoom on the Na{\,\sc i}5596--5590\AA\, region: the green
curve corresponds to the same model as above (blue) but with the sodium abundance
enhanced to four times cosmic.
{\bf Lower panel:} Montage of rectified spectra computed by including
bound-bound transitions of individual species, limited to those that affect the
emergent spectrum [color].
}
\label{fig_sn99em_14nov}
\end{figure*}

In the optical range, besides hydrogen lines, He{\,\sc i} features are numerous
with a strength of about one fifth of H$\alpha$ (see previous section
for a census of these lines).
However, the outflow has cooled down too significantly to show N{\,\sc ii} lines,
and we instead see the appearance of weak metal lines in the optical spectrum.
We have the Mg{\,\sc ii}(4f--3d) triplet at 4481\,\AA.
We also have Si{\,\sc ii}(5p--3d)4076.8, Si{\,\sc ii}(4f--3d)\,4128.0\AA, the Si{\,\sc ii}(4d--4p)
doublet at 5056.0\AA\, and the Si{\,\sc ii}(4p--4s) doublet at 6347.1\AA\
and 6371.4\AA.
We finally attribute some features to Fe{\,\sc ii} and Fe{\,\sc iii} but due to the large number of
contributors, we cannot enumerate them.
Note nonetheless that contrary to identifications by Leonard et al. (2002a),
the feature around 5200\AA\, has Fe{\,\sc iii} lines (groups of lines around 5080\AA\, and
5160\AA\,) as main contributors, rather than Fe{\,\sc ii}, and also overlaps with lines
of Si{\,\sc ii} at 5056\AA\, and He{\,\sc i} at 4921.9\AA\, and H$\beta$.
This careful log shows that it is difficult to associate ``features" in SN
spectra with {\it individual} lines of selected species.

We show in Fig.~\ref{fig_sn99em_5nov_ion} the outflow ionization for hydrogen, helium and iron for
the blue-curve model of Fig.~\ref{fig_sn99em_5nov}.
It resembles quite closely the previous case.
However, the He$^+$ region has shrunk, causing He{\,\sc i} lines to weaken.
Note also that H$^{0+}$ and Fe$^+$ have a larger fractional abundance in
the outer part.
The photosphere is located at $r= 1.59R_0$ ($\log(r/R_0) = 0.2$), where
H$^+$ and Fe$^{2+}$ dominate while helium is roughly equally present in its
first ionized and neutral forms.

\subsection{Late stage}
\label{Sec_late_stage}

We now turn to the description of a Type II SN spectrum, corresponding
to a late-stage in the photospheric-phase evolution.
In Fig.~\ref{fig_sn99em_14nov}, we show synthetic fits (color)
to the observations (black) of SN1999em taken on the 14th of November 1999.
The red curve corresponds to the continuum energy distribution of the model,
while the full synthetic spectrum is shown in blue.
The model parameters are:
$L_{\ast} = 1.5 \times 10^8 L_{\odot}$,
$R_{\rm phot} = 6.15 \times 10^{14}$ cm (or 8840$R_{\odot}$),
$v_{\rm phot}=6,350$\,km\,s$^{-1}$, $T_{\rm eff} =6,800$K, $n=10$,
$\rho_{\rm phot} = 8.7 \times 10^{-14}$ g\,cm$^{-3}$.

Only optical data is available for SN1999em at this date so we rely entirely on the
optical range to assess the UV flux level.
The absence of He{\,\sc i} lines and the increasing strength of Fe{\,\sc ii} lines are consistent
with a much cooler effective temperature compared to the previously discussed models,
associated with a strong line-blanketing in the UV range.
The lower panel shows that this line-blocking stems mostly from
Fe{\,\sc ii}, and to a lesser extent from Ni{\,\sc ii}, Ti{\,\sc ii} and Mg{\,\sc ii} (the 2800\AA\, feature).
We find that as long as a big model atom for Fe{\,\sc ii} is included in our
computations, the gross properties of the outflow and the emergent spectrum
remain unchanged when adding extra metal species, e.g. Ni{\,\sc ii}.

The optical range clearly shows the presence of P-Cygni profiles
from Ca{\,\sc ii}\,3933--3968\AA\, and Ca{\,\sc ii}\,8498--8542--8662\AA, which
gives further support for a relatively low ionization of the outflow.
Blends of H$\gamma$, Ti{\,\sc ii} and Fe{\,\sc ii} contribute to the feature
at 4300\AA\ while blends of Ti{\,\sc ii}, Mg{\,\sc ii} and Fe{\,\sc ii} contribute to the feature at
4600\AA.
Around 5000\AA\ we see a myriad of Fe{\,\sc ii} lines which
overlap with H$\beta$, while Fe{\,\sc iii} lines are no longer predicted.
The flux goes below the continuum level around 5500\AA, which we interpret
as the appearance of the doublet line Na{\,\sc i}\,5896--5890\AA.
Indeed, taking the same model and enhancing the sodium abundance by a factor
of four (Fig.~\ref{fig_sn99em_14nov}, top-panel, green curve in the inserted box), as expected
for a supergiant progenitor (Prantzos et al. 1986), gives a much better
fit to the observed sodium doublet.
The dip on the blue-side of this sodium doublet has been associated with
a number of Sc{\,\sc ii} lines (Leonard et al. 2002a),
a species not yet included in our atomic database.
Further to the red we reach a spectral region affected by weak Fe{\,\sc ii}
lines which, in combination with Si{\,\sc ii}6347--6371\AA\, gives rise to
some filling-in of the P-Cygni trough of H$\alpha$.

\begin{figure}[htp!]
\epsfig{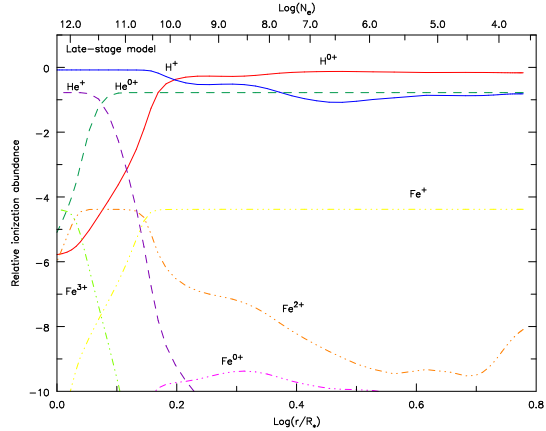}
\caption{
Relative ionization for hydrogen (solid), helium (dashed) and iron
(dash-dotted) for the late-stage model of Sect.~\ref{Sec_late_stage}.
Note that the highest ionization state of iron shown is Fe$^{3+}$ rather
than Fe$^{4+}$, but the lowest one is now Fe$^{0+}$ rather than Fe$^{+}$
[color].
}
\label{fig_sn99em_14nov_ion}
\end{figure}

As before, we show the outflow ionization predicted by the model
(Fig.~\ref{fig_sn99em_14nov_ion} -
note that the highest ionization state of iron shown is Fe$^{3+}$ but the
lowest one is now Fe$^{0+}$).
Unlike the previous two cases hydrogen recombines to its neutral state,
while in the corresponding region, iron is present in its Fe$^+$ state (giving
rise to the Fe{\,\sc ii} lines).
Note that in the regions where hydrogen recombination takes place,
the electron-density goes down more steeply with radius
(this is seen by inspection of the
spacing between tick marks on the top axis).
The photosphere is located at $r=1.48R_0$ ($\log(r/R_0) = 0.17$),
right in the recombination front: hydrogen (iron) is equally
represented in H$^+$ and H$^{0+}$ forms (Fe$^{2+}$ and Fe$^{+}$)
while helium is fully neutral.

\section{Discussion on model results and properties}
\label{Sec_mod_res}

\subsection{Hydrogen-recombination front in cool models}
\label{Sec_H_recom}

   For models having an ionization similar or lower to the case shown in Sect.~\ref{Sec_late_stage}
(late-stage model), we had to introduce an adaptive grid in CMFGEN in order to
resolve the region where hydrogen recombines.
Indeed, this tends to occur over very restricted spatial scales in the
outflow.

\begin{figure}[hbp!]
\epsfig{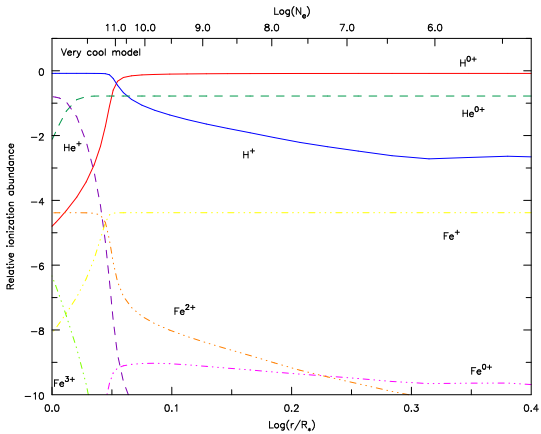}
\caption{
Relative ionization for hydrogen (solid), helium (dashed) and iron
(dash-dotted) for a very cool model.
For clarity, we only show the inner region of the outflow.
The radial variation of the electron-density is also shown on the
top axis [color].
}
\label{fig_ion_cool_mod}
\end{figure}

   We show in Fig.~\ref{fig_ion_cool_mod} a plot of the outflow
ionization for a very cool model with $T_{\rm eff} = 6000$K.
Note how steep the recombination of hydrogen is as well as the concomitant
recombination of iron from Fe$^{2+}$ to Fe$^+$.
Also shown is the radial variation of the electron-density (top axis);
tick marks are more closely packed around the front, following
the strong variation of the mean-electron number through this region.

We find that the shape of the front depends on the spatial scale of the
outflow, i.e. the bigger $R_0$ the more gradual the front is.
Further, for identical model parameters, a smaller turbulent velocity
gives reduced line-blanketing and thus a more extended hydrogen recombination front.
Reducing the hydrogen abundance also reduces somewhat the steepness of the front. This
may result from the fact that the electron-scattering optical depth is less
tied to the front itself, which sets the position of the photosphere for H-rich
models. Finally, varying the density exponent (see below) also modifies the shape
of the recombination front, by extending it over larger spatial scales for smaller
values of $n$.

\subsection{Choice of outer boundary}
\label{Sec_out_bound}

There is a strong computational incentive to limit the extent of the simulated
grid.
Ideally, the outer grid location should be where the medium becomes
optically thin at all wavelengths.
Unfortunately, H{\,\sc i} and He{\,\sc i} bound-free edges in the UV, as well as a
few resonance lines remain optically thick even far above the photosphere.
This occurs even though the density at r is only $(R_0/R_{\rm max})^n$ its value
at $R_0$, $n$ being of the order of 10.
The drawback of stretching the computation to large radii is that the maximum velocity
in the outflow can become very large, i.e. up to one third of the speed of light
for the modeling, e.g., of the spectrum of SN1987A on the 24th of February 1987.
In expanding outflows, a line with frequency $\nu_0$ at a given wind location $r$
can in principle interact with photons emitted at $r'$ with a frequency between $\nu_0$ and
$\nu_0 (1 + {\bf n} \cdot ({\bf v(r)} - {\bf v(r)'}))$ where {\bf n} is the unit vector
linking the two points of interest.
The frequency sampling around line center must therefore be fine enough to account
for potential line-transfer effects.
But the larger the maximum velocity in the computation is, the larger is the frequency array
and thus the more computationally demanding the model is.

\begin{figure*}[htp!]
\epsfig{file=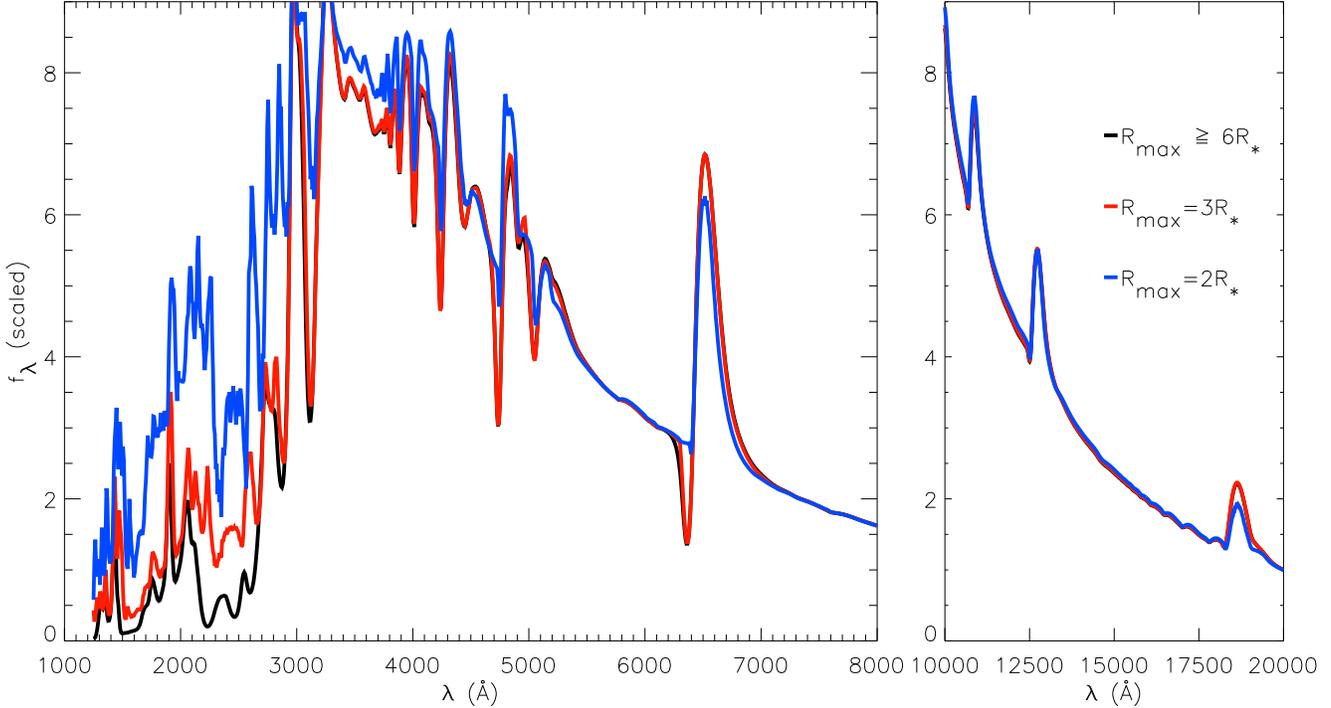, width=18cm}
\caption{
Comparison between emergent spectra computed with identical parameters
but differing in the adopted maximum radius, chosen at 2 (blue), 3 (red)
and 6, 10 and 15$R_0$ (overlapping perfectly to form the black curve).
To facilitate the comparison of the relative synthetic fluxes for the different models,
these have all been normalised to unity, at 10,000\AA\, (left panel)
and 20,000\AA\, (right panel).
$R_0$ is identical in all cases.
Note also that the red and black curve overlap perfectly everywhere
except in the UV.
We employ a small model atom for this investigation, i.e. solely H{\,\sc i},
He{\,\sc i}, C{\,\sc ii}, N{\,\sc i}, O{\,\sc i}, Fe{\,\sc ii}, Fe{\,\sc iii}
[color].
}
\label{fig_max_rad}
\end{figure*}

In general, our simulations extend out to 10$R_0$.
But we have investigated whether the choice of a bigger or a smaller
maximum radius would produce a noticeable effect on the SED.
Models suitable for the earlier evolution of Type II SN
during the photospheric phase (first week) tend to require
a higher density exponent than at later times (see, e.g., H\"{o}flich 1988
for SN1987A; Eastman \& Kirshner 1989), minimising the effect
resulting from the modulation of the location of the maximum radius.
For this check, we therefore have chosen a model with a density exponent
of 8.
This study is done with a small model atom, including only the most
abundant species H{\,\sc i}, He{\,\sc i}, C{\,\sc ii}, N{\,\sc i}, O{\,\sc i},
Fe{\,\sc ii}, and Fe{\,\sc iii}.

We show in Fig.~\ref{fig_max_rad}\ the SED for models with a maximum radius of 2, 3, 6, 10,
and 15$R_0$.
For this latter case, we have modified the Hubble law by forcing it to
smoothly plateau beyond 10$R_0$ to reach at 15$R_0$ only 110\% of the value
it has at 10$R_0$.
Only three curves are seen in Fig.~\ref{fig_max_rad}.
Indeed, we find that the SED for the cases where $R_{\rm max} \ge 6 R_0$
are indistinguishable (they overlap under the black curve), a very desirable
property which therefore justifies a significant reduction of the maximum
radius.
Out of curiosity, we have also reduced the maximum radius down to 3 (red
curve) and finally 2$R_0$ (blue curve).
Interestingly, beyond 3000\AA, the
spectrum computed with $R_{\rm max} = 3 R_0$,
is indistinguishable from that computed with $R_{\rm max} = 6 R_0$.
This suggests that the continuum and line formation beyond 3000\AA\, is
confined to the photosphere, located in these models at 1.8$R_0$.
However, the UV flux lies significantly above the curve corresponding
to the model with $R_{\rm max} = 6 R_0$, which results from
the neglect of line-blocking from the outflow layers between 3 and 6$R_0$.
Indeed, metal lines in the UV remain optically thick far above the
photosphere.
Going down to $R_{\rm max} = 2 R_0$, we see that the UV flux is even more
in excess, for the same reason given as above, only magnified by the further
reduction in maximum radius.
But additionally, hydrogen lines (most notably H$\alpha$, H$\beta$,
H$\gamma$, P$\alpha$) are affected, both in emission and
absorption. Note that the latter is nearly absent in H$\alpha$,
suggesting its P-Cygni trough forms over an extended region.
In fact, adopting such a small value for $R_{\rm max}$ has
a strong influence on the non-LTE solution of the problem.
For example, we now find that the outer radius is optically
thick throughout the wavelength range, so that there is
no photosphere in this model!

Although the tests with $R_{\rm max}$ of 2 and 3$R_0$ presented
only pedagogical material, the adequacy of choosing $R_{\rm max}= 6 R_0$
in our simulations is a very useful result.

   \subsection{Influence of metallicity on the emergent spectrum}
\label{Sec_inf_met}

\begin{figure*}[htp!]
\epsfig{file=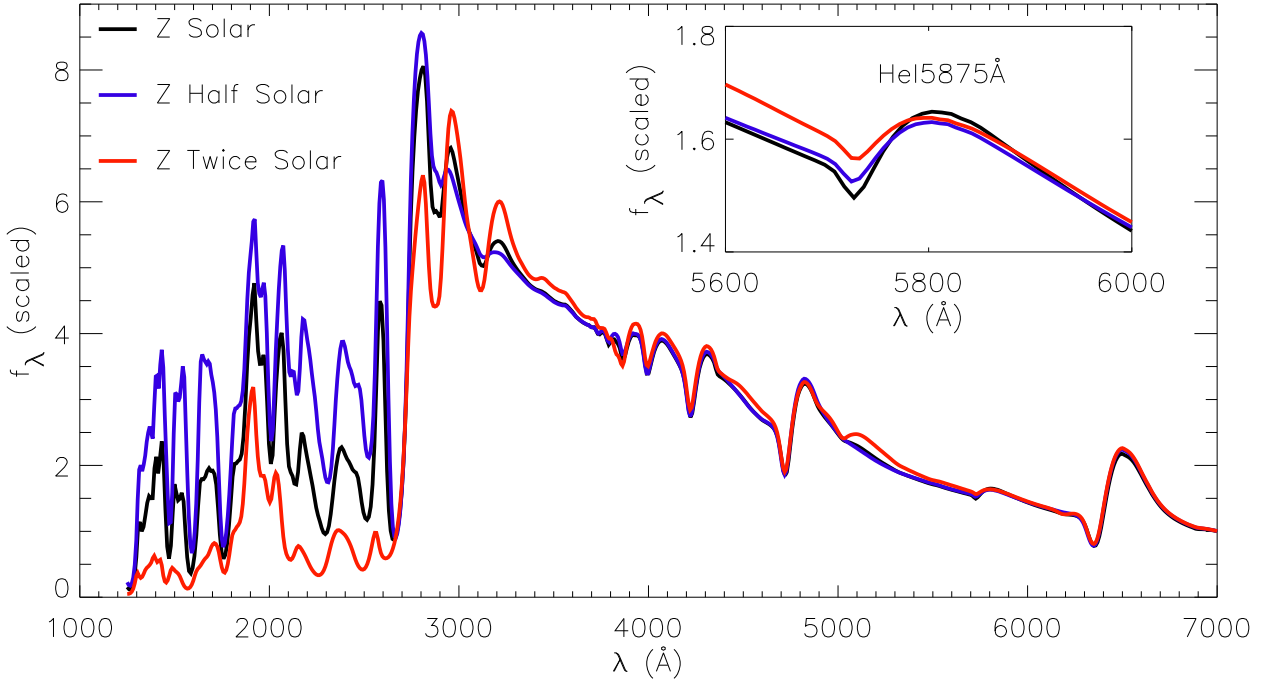, width=18cm}
\caption{
Comparison of the emergent flux from models with identical parameters apart
from a different adopted
metallicity. The black curve describes a model with solar metal abundance.
The red (blue) shows the resulting emergent spectra when all metal abundances
are scaled by a factor of two (one half).
The model spectra beyond H$\alpha$ are relatively insensitive to changes in
metallicity --- therefore for ease of comparison we have normalised the fluxes
to unity at 7000\AA\, [color].
}
\label{fig_inf_met}
\end{figure*}

As was shown in Sect.~\ref{Sec_spec_mod}, line-blanketing becomes very effective in blocking the
continuum star light as soon as the medium has cooled down sufficiently for Fe{\,\sc ii} to
be present.
The effective temperature at which this transition occurs is around 8,000K.
Starting from the intermediate-stage model of Sect.~\ref{Sec_int_stage}, we have investigated what
the impact would be of reducing or enhancing the metallicity, i.e scaling all metal
abundances by a factor of a half or two.
We show the results in Fig.~\ref{fig_inf_met}, where we reproduce the model of
Sect.~\ref{Sec_int_stage}\ in black,
the twice-solar (half-solar) metallicity model in red (blue).
As expected, the higher the metallicity, the more significant the line-blocking, and this
is seen most clearly in the reduction of the emergent flux in the UV.
Going from the model with $Z=0.5 Z_{\odot}$, to $Z=Z_{\odot}$ and $Z=2 Z_{\odot}$,
the Rosseland-mean optical depth at the base of each model is increased by about 10\%,
the photosphere moves outwards by less than 1\% of the radius in the thermalisation layer,
with a decrease in effective temperature and an increase in the velocity of the same order.
Hence, despite the fact that each model possesses essentially the same outflow properties,
the different adopted metallicity leads to significant differences in the UV spectral
morphology.
The optical spectral morphology and in particular the slope of the SED is essentially
unchanged, reflecting the similarity of each model properties.
The only differences are a somewhat weaker He{\,\sc i}\,5875\AA\, line at higher
metallicity, perhaps resulting from the reduced strength of the ionizing flux,
and an enhanced strength of Fe{\,\sc ii}/Fe{\,\sc iii} lines around 5200\AA\, and 4500\AA.

It thus seems that the UV SED can be used to constrain the environmental metallicity
at early times when the line-blocking due to the Fe{\,\sc ii} lines is still moderate.
At later times, e.g., for the late-stage model discussed in Sect.~\ref{Sec_late_stage},
the UV flux is negligible and thus placing constraints on the metallicity has to be 
done from the optical - there are no
strong iron lines at longer wavelengths. Strong line-overlap makes this task difficult.
Finally, note that the probability of seeing metals nucleosynthesized during
the explosion increases with time so the primordial metallicity of
the progenitor can be more reliably constrained from
spectra taken at earlier times.


\subsection{Influence of Luminosity}
\label{Sec_inf_lum}

\begin{figure*}[htp!]
\epsfig{file=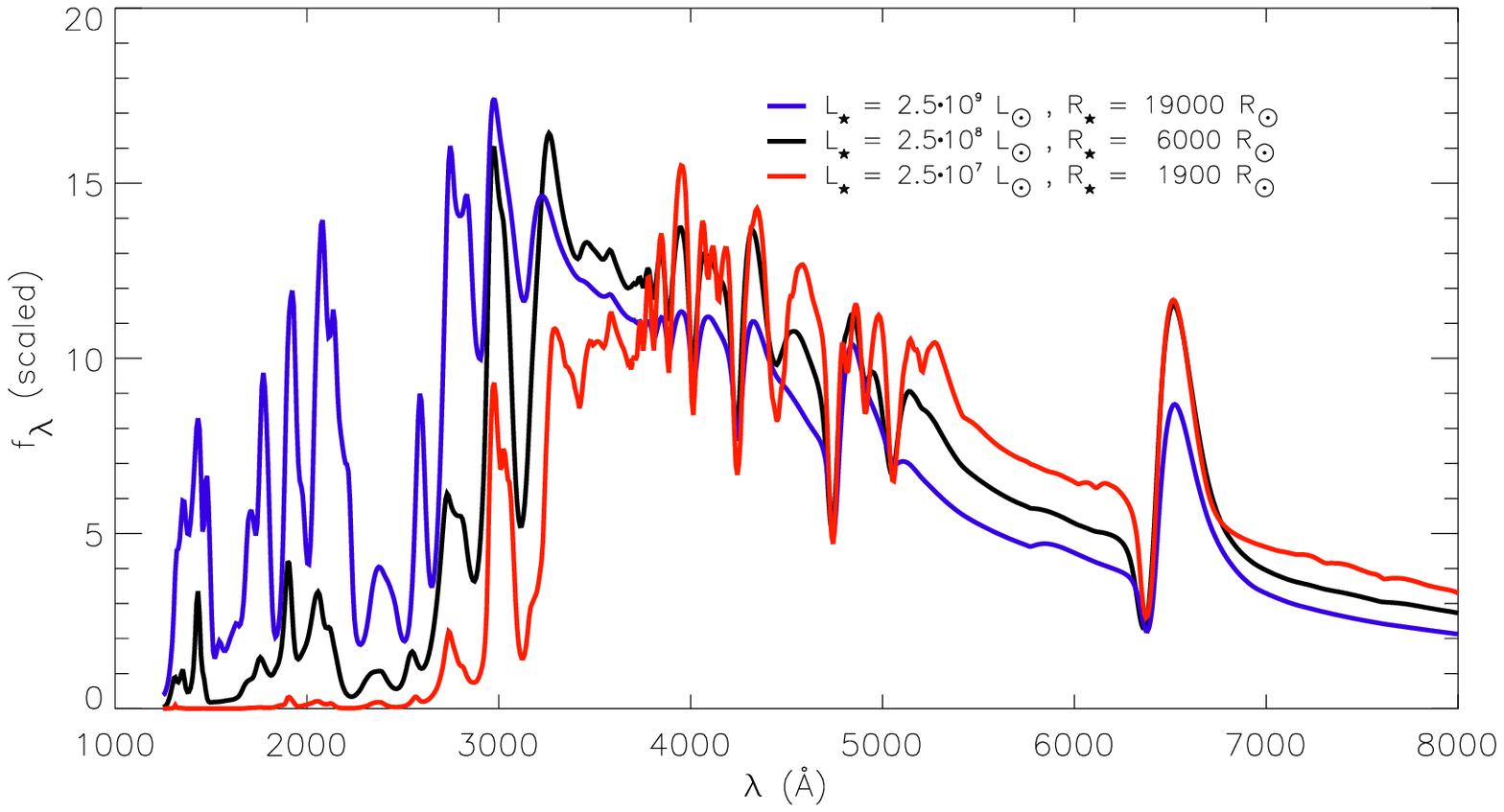, width=18cm}
\caption{
Comparison of the emergent flux from models differing in luminosity, enhanced
and decreased by a factor of ten compared to the model in black.
Also, we scale the radius and the base density up or down by $\sqrt{10}$, to maintain
the same photospheric velocity for all (see text for details).
The model atom used is small, including only the most abundant species, i.e. H{\,\sc i},
He{\,\sc i}, C{\,\sc ii}, N{\,\sc i}, O{\,\sc i}, Fe{\,\sc ii} and Fe{\,\sc iii}
[color].
}
\label{fig_inf_lum}
\end{figure*}

When modeling SN there is always an uncertainty in its distance. In this
section we therefore investigate how the synthetic flux distribution changes 
with luminosity.
Note that in our current approach the luminosity of the model is adjusted
so that, for an adopted Cepheid distance,
the model flux distribution lies within a factor of 2 of that observed.
In the future, it will be desirable to modify CMFGEN in order to use
a SN luminosity computed from a hydrodynamical simulation of the explosion.

We show in Fig.~\ref{fig_inf_lum}\ a set of three models differing in model luminosity
(black curve: $L_{\ast} = 2.5 \times 10^8 L_{\odot}$;
blue curve: $L_{\ast} = 2.5 \times 10^9 L_{\odot}$;
red curve: $L_{\ast} = 2.5 \times 10^7 L_{\odot}$).
In order to maintain a similar total base continuum optical depth
and effective temperature, we scale up (down) the base radius
by $\sqrt{10}$ and scale down (up) the base density for the model corresponding
to the blue (red) curve.
To position the three models in the same figure, we display the flux of the
blue (red) model scaled down (up) by a factor of 10.
With this scaling the model that has a higher flux in the UV has a lower flux
in the optical and vice-versa, a likely result of the modulation of line-blanketing
due to the variation in spatial scale of the corresponding model.
Note that we use a small model atom, including only the most
abundant species H{\,\sc i}, He{\,\sc i}, C{\,\sc ii}, N{\,\sc i}, O{\,\sc i},
Fe{\,\sc ii} and Fe{\,\sc iii}.

These various model scalings lead to a very similar photospheric velocity, i.e.
6,700\,km\,s$^{-1}$ (red curve) and 7,200\,km\,s$^{-1}$ for both black and blue curves,
as can be seen from the nearly overlapping positions of absorption in line features.
As expected from Fig.~\ref{fig_inf_lum}, the effective model temperature increases from the red
($T_{\rm eff} = 7,000$K), to the black ($T_{\rm eff} = 7,900$K) and blue model
($T_{\rm eff} = 8,400$K).
The reason for this is that the photospheric density varies a great deal
amongst those three models, increasing from high to low luminosity models
(blue: $\rho_{\rm phot} = 0.88 \times$ 10$^{14}$\,g\,cm$^{-3}$;
black: $\rho_{\rm phot} = 2.8 \times$ 10$^{14}$\,g\,cm$^{-3}$;
red: $\rho_{\rm phot} = 16.3 \times$ 10$^{14}$\,g\,cm$^{-3}$).
The higher the photospheric density the stronger the line-blanketing and
the more depleted the UV flux appears.

Thus, when a large flux difference between a model and a given observation
exists, a satisfactory fit to the observed energy distribution will likely
become poor once the model and observed flux levels are comparable.
We find that little or no changes are observable in the synthetic flux
distribution when the luminosity is varied by a factor of two, so this
sets the error one can make on the model luminosity when performing
fits to observations.

\subsection{Effects of varying the density exponent}
\label{Sec_den_exp}

In our approach, we use an analytical description of the outflow density,
characterized by a base density $\rho_{0,n}$ and a density exponent $n$ so that
at a given height $r$, we have $\rho(r) = \rho_{0,n} (R_{\ast}/r)^n$.
To ensure that the diffusion approximation is fulfilled at the base of the envelope,
the base density is chosen so that the total continuum optical
depth of the outflow is approximately 50.
The density exponent is, however, constrained from observations.
We have therefore a lot more flexibility than if we based our calculations
on hydrodynamical inputs, for which the density distribution is set once
and for all at the start of the homologous expansion.
This flexibility is not necessarily a benefit since
varying the density exponent can influence the emergent spectrum in a variety of
ways, widening considerably the already large parameter space.

\begin{figure*}[htp!]
\epsfig{file=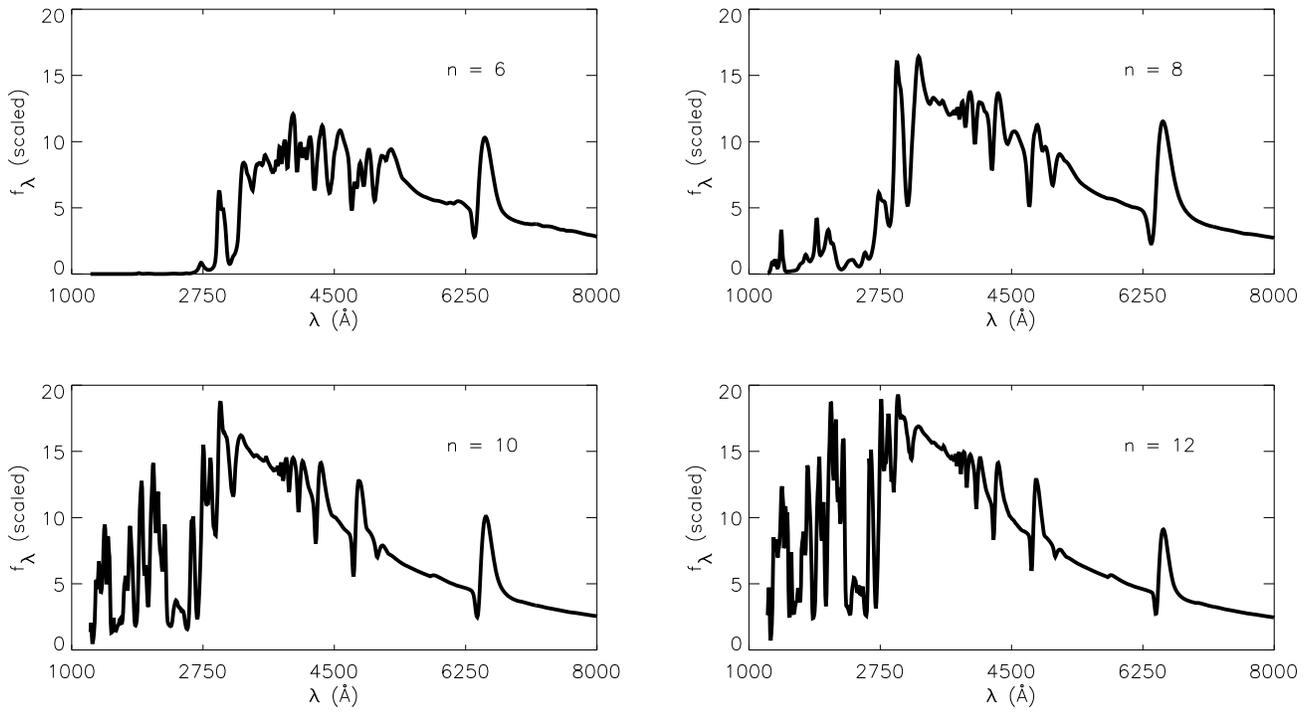, width=18cm}
\caption{
Illustration of the influence of the density exponent $n$ on the synthetic flux distribution
for $n=6, 8, 10$ and $12$. All other model parameters are kept identical.
The model atom used is small, including only the most abundant species, i.e. H{\,\sc i},
He{\,\sc i}, C{\,\sc ii}, N{\,\sc i}, O{\,\sc i}, Fe{\,\sc ii} and Fe{\,\sc iii}.
}
\label{fig_inf_den}
\end{figure*}

\begin{figure*}[htp!]
\epsfig{file=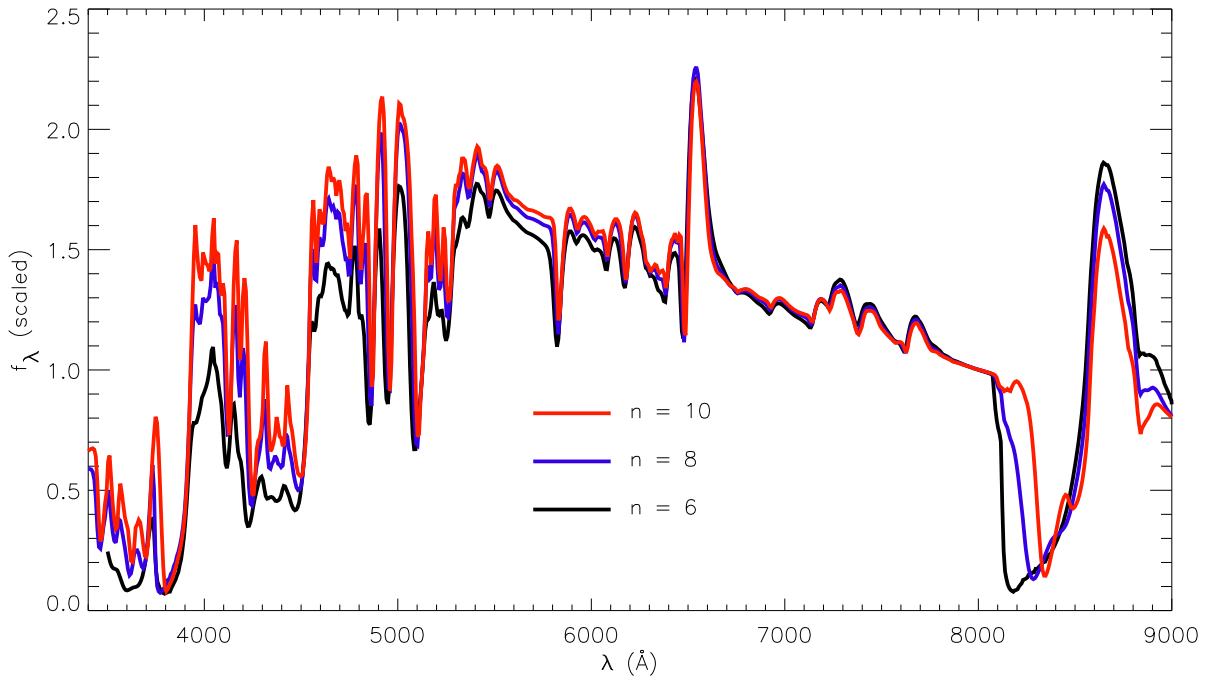, width=18cm}
\caption{
Illustration of the influence of the density exponent $n$ on the synthetic flux distribution
for $n=6, 8, 10$ in a cool model where hydrogen recombines near the base.
All other model parameters are kept identical.
Note that the most striking effect is on the set of Ca{\,\sc ii} lines near
8500\AA, the line appearing broader and stronger for smaller density
exponents [color].
}
\label{fig_inf_den_cm}
\end{figure*}

In the previous section, we saw that the spatial scale of the expanding SN
can affect significantly the emergent spectrum, mostly from the variation
of the photospheric density.
Let us assume that the model is fully ionized and that the continuum opacity
is mostly due to electron-scattering ($\kappa_e \sim 0.34$ cm$^2$/g),
and investigate how model properties are affected by a change in
the characteristics of the density distribution.
Along a radial ray, the continuum optical-depth is given by
$\tau(r) = \int_r^{\infty} \kappa_e \rho(r') dr' $ so that at the photosphere,
we
have $
\tau_{\rm phot} = \frac{\kappa_e \rho_{0,n} R_{\ast}}{n-1}  \left( \frac{R_{\ast}}{R_{{\rm phot},n}} \right)^{n-1}$

Let us take another model identical in all parameters except for those that characterize the
density distribution ($n'$, $\rho_{0,n'}$, $R_{{\rm phot},n'}$)
and compare these at the photosphere:
$$
 \frac{\rho_{0,n} R_{\ast}}{n-1}  \left( \frac{R_{\ast}}{R_{{\rm phot},n}} \right)^{n-1} =
 \frac{\rho_{0,n'} R_{\ast}}{n'-1}  \left( \frac{R_{\ast}}{R_{{\rm phot},n'}} \right)^{n'-1}
$$

Within this simple approach we can determine how the photospheric radius changes
if we keep the base density (and $R_{\ast}$) the same for both models,
i.e. $\rho_{0,n} = \rho_{0,n'}$.
We obtain
$$
  \frac{R_{{\rm phot},n'}}{R_{\ast}} = \left(  \frac{n-1}{n'-1}
\right)^\frac{1}{n'-1}
\left( \frac{R_{{\rm phot},n}}{R_{\ast}}  \right)^\frac{n-1}{n'-1}
$$
Thus, for identical base density, if $n' > n$, $R_{{\rm phot},n'} < R_{{\rm phot},n}$,
i.e. the photosphere moves in to hotter and slower regions.

We show in Fig.~\ref{fig_inf_den}\
the synthetic spectra of models differing solely in density exponent $n$ - in
particular the base density is kept constant - covering values of 6, 8, 10 and 12.
We clearly see that as the exponent is increased, the spectrum shows a more and more
pronounced UV flux, metal lines of decreasing strength in the optical but a strengthening
He{\,\sc i}\,5875\AA. As expected, the photospheric radius decreases from 1.9, to 1.81, 1.54 and 1.4
times the base radius (kept constant) as one goes from $n=6$ to $12$.
Given the adopted Hubble law, a similar decrease in photospheric velocity follows (from
7,600 to 7,250, 6,150 and 5,600\,km\,s$^{-1}$).
But there is a significant rise of the effective temperature over that range,
from 6,400, to 7,900, 8,100 and 8,250K.
These effects are reversed if instead one maintains the photospheric radius
rather than the base density when changing the density exponent
(Eastman \& Kirshner 1989).

\begin{figure}[htp!]
\epsfig{file=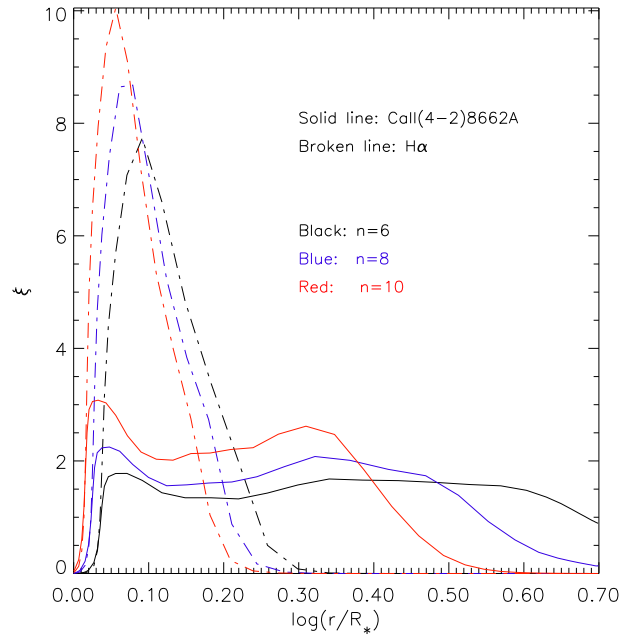, width=9cm}
\caption{
Illustration of the relative line emission occurring at a given height in the outflow, both
for one of the three components of the Ca{\,\sc ii} feature at ca. 8500\AA,
i.e. Ca{\,\sc ii}8662\AA\, (solid line) and H$\alpha$ (broken line).
We use a color coding to differentiate between models with different density exponent.
Note how extended the line formation region of the Ca{\,\sc ii} line is compared to
that of H$\alpha$.
The other two components of the Ca{\,\sc ii} triplet show a similar behavior and
are thus not shown.
$\xi$ is defined such that line flux is proportional to $\int \xi d\log r$ [color].
}
\label{fig_orig_caii}
\end{figure}

Besides the general influence on the spectral energy distribution,
changing the density exponent has also a significant effect on lines.
For example, there is a systematic blue-shift of the observed H$\alpha$
P-Cygni profile in Type II SN, which can be modulated in models by changing the
density exponent (Sect.~\ref{Sec_line_form}).
Here we illustrate the influence the density exponent has on the strength and width
of the Ca{\,\sc ii} feature near 8500\AA.
This set of lines is present only in models that are cool enough, i.e. those
where hydrogen recombines at the photosphere.
We show in Fig.~\ref{fig_inf_den_cm}\ the synthetic flux distribution of three models
differing only in density exponent ($n=6, 8$ and $10$).
Contrary to the previous figure, the effect on the overall spectrum is small,
mostly because in this parameter space, the line and continuum formation is essentially
contained within the (localized) region of the recombination front.
However, as shown in Fig.~\ref{fig_orig_caii}, the Ca{\,\sc ii} lines at 8498, 8542 and 8662\AA\,
form over the entire outflow and are therefore extremely sensitive to the density gradient even
above the recombination front, resulting in a broader and stronger Ca{\,\sc ii} feature for smaller exponents.
However, H$\alpha$ is more confined to the photosphere region and is thus only weakly sensitive to the
outer density distribution.
This explains why this set of Ca{\,\sc ii} lines can be used to reveal the presence of outflow
asymmetry, as inferred, e.g., by Kasen et al. (2003) in the Type Ia SN2001el, from the presence
of strong polarization (with a different angle) in this Ca{\,\sc ii} spectral region
compared to the rest of the spectrum.
This line can therefore be used as a diagnostic for the density distribution
and/or departure from sphericity of SN outflows in general, when ionization conditions
in the outflow are suitable, i.e. for effective temperatures at and below 8000K.

\subsection{Lines of selected species}
\label{Sec_sel_lines}

   \subsubsection{Hydrogen lines}

At present there is an open debate concerning the capability
of model atmospheres to fit the hydrogen Balmer lines of
Type II SN during the photospheric phase.
Some groups fail to reproduce these lines (Schmutz et al. 1990
and Mitchell et al. 2002, both for SN1987A) while others fit them
very well (Eastman \& Kirshner 1989; H\"{o}flich 1988, again for SN1987A).
Schmutz et al. invoked the possibility of clumping although they did not
investigate whether it could in reality cure the problem, while a decade
later, Mitchell et al. invoked energy deposition following $^{56}$Ni mixing
in the outer envelope of SN1987A, as soon as 4 days after core-collapse.
Those who fit the hydrogen Balmer lines well for the same dates do
not use clumping and ignore energy deposition.

This situation is unsatisfactory since it is difficult to establish whether
the problem is physical or numerical.
We have shown in the previous section that CMFGEN reproduces extremely well
the hydrogen Balmer lines, even though we {\it overplot} the synthetic
spectrum and adopt a {\it linear flux scale}, two elements that have the
potential to make fit discrepancies very clear.
Although we show in Sect.~\ref{Sec_spec_mod} only three cases, all synthetic spectra
computed for SN1987A and SN1999em fit the observations very well,
provided that a strong hydrogen recombination front does not develop,
i.e. as long as hydrogen remains mostly ionized in the outflow.
Indeed, if such a front appeared, CMFGEN always predicted H$\alpha$ too weak
by a factor of 2-3 compared to observations, both for the emission and absorption
components (see for example the H$\alpha$ region in the cool models shown in
Fig.~\ref{fig_inf_den_cm}).


In CMFGEN, whose principal purpose is to determine
the atmospheric structure and level populations, we simply adopt a Gaussian shape,
of constant width, to describe the intrinsic profile. In practice,
the width of the profiles is set by the adopted micro-turbulent velocity.

To save computational time, we have chosen a turbulent velocity $v_{\rm turb}$
of 100\,km\,s$^{-1}$.
For fully ionized models, varying this value by a factor 2-3 has
no effect on the emergent spectrum.
However, for cooler models, we find that reducing this turbulence to 20\,km\,s$^{-1}$
is important in order to reproduce better the profile shape and strength of
H$\alpha$. In those cool models, a lower turbulent velocity
gives a SED with a higher flux in the UV, and
essentially a higher apparent outflow ionization (in fully ionized models,
the effect on the SED is unnoticeable).
We understand this as resulting from the reduced line-blanketing,
itself stemming from the reduced spectral band-width over which metal lines
block the ionizing flux emanating from the thermalisation layer at the
base.
Similarly, late-stage simulations performed with fewer metal species
show an appreciable increase in the H$\alpha$ strength, again likely to
result from the modulated amount of line-blanketing.


   \subsubsection{Helium lines}
\label{Sec_hei}
We have seen in Sect.~\ref{Sec_inf_met}  (Fig.~\ref{fig_inf_met})
that varying the metallicity, and hence the magnitude of
the line-blanketing, changes the strength of the He{\,\sc i}\,5875\AA\, line.
This He{\,\sc i} line has proven to be difficult to model, even impossible without
invoking unreasonable enrichments in helium (Schmutz et al. 1990,
Eastman \& Kirshner 1989).
However, as can be seen from the detailed computations of Mitchell et al. (2002)
for SN1987A and Baron et al. (2000) for SN1999em, PHOENIX reproduces very well
this feature in the early spectra taken.
This seems to suggest that treating in non-LTE the metal lines that are responsible
for the line-blocking is important.

\begin{figure}[htp!]
\epsfig{file=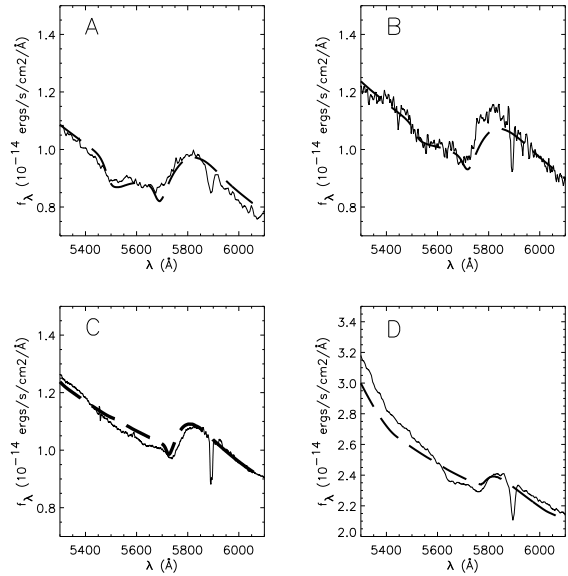, width=8cm}
\caption{
Synthetic fits (broken line) to observations (solid line) of SN1999em
taken on the 30th of October (A), and the 1st (B), 5th (C) and 8th (D)
of November 1999 (Leonard et al. 2002a), showing the
He{\,\sc i}\,5875\AA\, line  region.
Note how well CMFGEN manages to reproduce the observed profile using
only a modest helium enrichment over the cosmic value (H/He=5 by number),
in agreement with stellar evolutionary predictions for Blue/Red 
supergiants, progenitors of Type II SN.
This capability of CMFGEN is in stark contrast with failures to reproduce this
He{\,\sc i} line in the works of Schmutz et al. (1990) and Eastman \& Kirshner (1989),
suggesting the importance of treating all species in non-LTE.
}
\label{fig_hei}
\end{figure}

We show in Fig.~\ref{fig_hei} CMFGEN fits to He{\,\sc i}\,5875\AA\, observed
in SN1999em on the 30th of October
and on the 1st, 5th and 8th of November 1999 (Leonard et al. 2002a).
The quality of the fits is very good, with only a modest helium enrichment (H/He $=$ 5 by
number).

   \subsubsection{Nitrogen lines}
\label{Sec_nit}

Based on a direct analysis with SYNOW, Baron et al. (2000) proposed that the
feature blue-ward of He{\,\sc i} 5875\AA\, in the spectrum of SN1999em taken on
the 30th of October was due to N{\,\sc ii}.
However, they could not confirm this statement with the more detailed
analysis done with PHOENIX.
They also highlighted the presence of a kink in the blue wing of
H$\beta$, which they do not associate with N{\,\sc ii}, but rather to some
peculiar emission process occurring in the outflow.
Similar intriguing features were observed in the early spectra of another
Type II SN, SN1999gi by Leonard et al. (2002b).
They observed the presence of un-identified features in the blue wing of
both He{\,\sc i}\,5875\AA\, and H$\beta$, at blue-shifted velocities of ca. 20,000 to
30,000\,km\,s$^{-1}$ although, these have now clearly-defined P-Cygni profile shapes.
Given the relatively higher density exponent characterizing the outflows of Type II
SN at earlier dates, it seems unlikely that much emission could arise at
velocities (or distances), a few times the value at the photosphere,
especially in recombination lines since these have a density-square
dependent emissivity.

\begin{figure}[htp!]
\epsfig{file=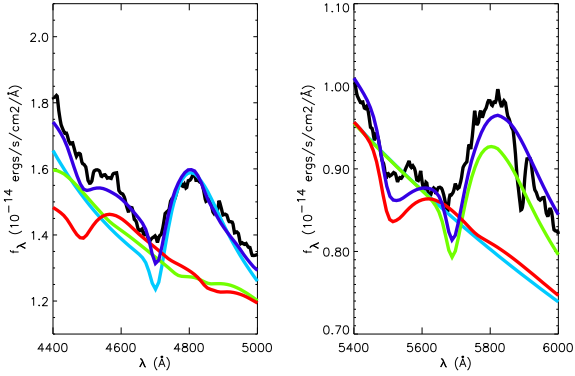, width=8cm}
\caption{
Zoom into the optical spectroscopic observations (black) of SN1999em taken on
the 30th of October
1999 (Leonard et al. 2002a) showing synthetic fits for a model comprising all
species
as given in Table 1 (blue) or only bound-bound transitions of hydrogen
(turquoise),
helium (green) and nitrogen (red).
We strongly support the idea that the observed features around 4600\AA\,
and 5500\AA\, are due to N{\,\sc ii} lines [color].
}
\label{fig_nit}
\end{figure}

We show in Fig.~\ref{fig_nit} synthetic fits to observations of
SN1999em taken on the 30th  of October 1999 (Leonard et al. 2002a).
The dark-blue curve corresponds to the emergent spectrum when all
species are included.
It fits well the He{\,\sc i}\,5875\AA\, and H$\beta$ regions, although
a slightly higher expansion velocity would improve the fit to the
He{\,\sc i} line (i.e., a shift of the line formation region to larger
velocities).
Note the presence of kinks at around 4600\AA\, and 5500\AA, i.e. in
the blueward region of each of these two lines.
We over-plot the emergent spectra with bound-bound transitions
exclusively of hydrogen (turquoise), helium (green) and nitrogen
(red curve), normalizing all curves to the observed flux at 8000\AA.
Unlike PHOENIX, CMFGEN is capable of reproducing the N{\,\sc ii}
line in the blue wing of He{\,\sc i}\,5875\AA, but together with
this line, it also predicts the equally strong N{\,\sc ii} feature
around 4600\AA.
In our model, these correspond to multiplets of N{\,\sc ii}(3d--3p)
centered around 4480\AA\, and 5490\AA, as well as N{\,\sc ii}(3p--3s)
at 4630\AA\, and 5680\AA.
 
The reason for the different results between PHOENIX and CMFGEN, regarding the
N{\,\sc ii} lines and the kink in the blue wing of H$\beta$, is unclear.
Our fits to hydrogen Balmer line profiles are much better, in particular the 
width and depth of the troughs, giving us confidence that we are not predicting 
spurious absorption too far to the red of line centers.
Moreover, their Fig. 6 suggests that not just H$\beta$ shows a kink beyond the
point where the absorption trough meets the continuum, but that this feature
is also present at a similar location for He{\,\sc i}5875\AA\, and H$\alpha$
(note that their use of a very thick line makes this difficult to see).
Such a kink is however {\it not} observed in H$\alpha$ (see their Fig. 3).


Our model fits show unambiguously that nitrogen lines are present in the optical spectrum
of the early photospheric phase of SN1999em.
In SN1999gi, spectroscopic observations taken after the 12th of December 1999
are very similar to those of SN1999em, so that we reach the same conclusion.
Note however that for the first spectrum of SN1999gi (taken on the
10th of December 1999), the feature at ca. 4600\AA\, is much stronger than
that at ca. 5500\AA, a situation we have not been able to reproduce.
By the 12th, the feature has both receded and weakened and is then
well reproduced (and identified as an N{\,\sc ii} line) by our model for SN1999em
on the 30th of October 1999 (Fig.~\ref{fig_nit}).
In our model, we use a factor ten enrichment of nitrogen compared to cosmic
while we find that these N{\,\sc ii} features become unnoticeable when this factor
becomes smaller than around 3.
This strongly supports the idea that the progenitors of SN1999em and SN1999gi
had core-processed nitrogen at their surface at the time of collapse.

Interestingly, although the first spectrum of SN1987A shows the high
ionization conditions for the presence of these N{\,\sc ii} features, these
are not observed.
Together with the fact that a modest helium enhancement permits a nice
fit of the spectrum, this suggests that the pre-collapse progenitor of
SN1987A had a surface abundance closer to the current LMC abundances.
Hence, identification of N{\,\sc ii} lines is a significant clue for
enhanced nitrogen abundance in the outer layers of the progenitor star.
Note that the over-abundance of nitrogen should be accompanied by
a significant depletion of carbon and oxygen if it resulted mostly from
the peeling-off of the outer envelope through a stellar wind mass loss.
In the case of mixing, which can for example be made more efficient through
stellar rotation, a more modest depletion in carbon and oxygen
would result. Unfortunately, we have not been able to identify lines from
either of these species in the photospheric phase of Type II SN,
although the addition of C{\,\sc i} in our model atom could alter this conclusion.
We defer until a future study a proper discussion on this issue.

\section{Line formation in expanding outflows of
hot stars and Type II SN}
\label{Sec_line_form}

Before concluding on these premiminary results with CMFGEN, we
present a study on the formation of P-Cygni line profiles in Type II SN outflows,
which reveals subtle differences from that observed, e.g., in the eponymous star.
Indeed, a striking feature in early-time spectra of Type II SN is the large
blue-shift, by as much as ca. 100--150\AA\, (5000--7000\,km\,s$^{-1}$),
of the H$\alpha$ line profile compared to the rest wavelength of 6562.79\AA.
At later times, this blue-shift is less obvious, partially because the
photospheric velocity quickly decreases, but it is nonetheless
present (see, e.g., Pastorello 2004).

To investigate the origin of this feature, we first focus on the formation
of scattering (resonance) and recombination lines in Wolf-Rayet outflows.
Their density distribution is much flatter than that of SN, following roughly a $1/r^2$ law
(imposed by mass conservation); non-LTE effects are
important but line-formation occurs over much larger spatial scales
compared to the photospheric radius.
Therefore, these objects offer a nice counterpart to line-formation conditions
in SN, which we are trying to understand.

\begin{figure}[htp!]
\epsfig{file=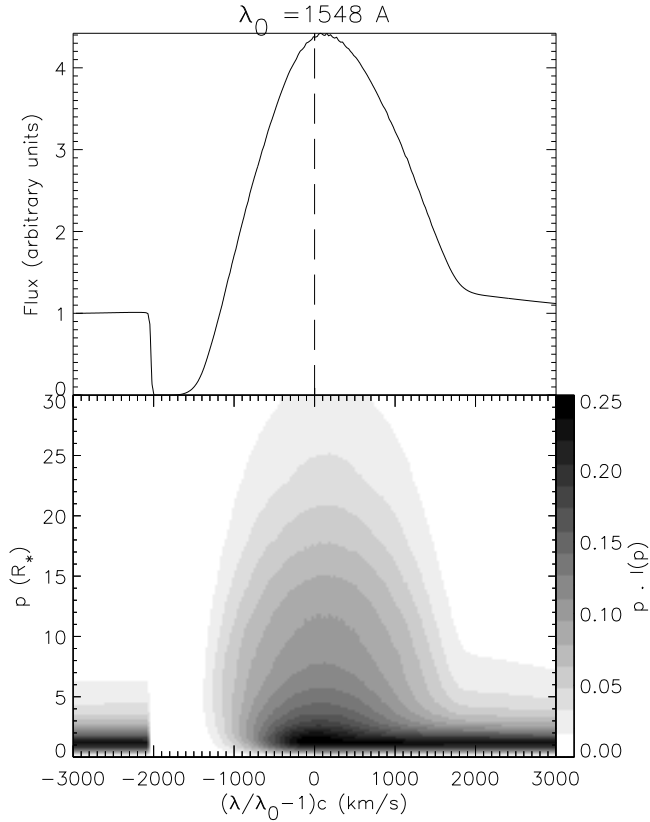, width=8cm}
\caption{{\bf Hot star model (WN5).} {\bf Bottom:} Grayscale image of the
quantity $p \cdot I(p)$ as a function of $p$
and scaled wavelength $x=(\lambda/\lambda_0-1)c$, where $p$ is the impact parameter
(in units of the hydrostatic radius R$_{\ast}$), $I(p)$ the specific intensity along
$p$ (at $x$). Here, $\lambda_0$ corresponds to the rest wavelength of C{\,\sc iv}\,1548\AA\, and $c$
is the speed of light. The terminal velocity of the model is 2000\,km\,s$^{-1}$.
{\bf Top:} Line profile flux, directly obtained by summing $p \cdot I(p)$ over the range of $p$.
Thus, the line flux at $x$ in the top panel corresponds to the cumulative sum of all contributions
$p \cdot I(p)$ at $x$ shown in the bottom panel, giving a vivid illustration of the sites
at the origin of the observed line profile.
}
\label{fig_jnu2_wn5}
\end{figure}

\begin{figure}[htp!]
\epsfig{file=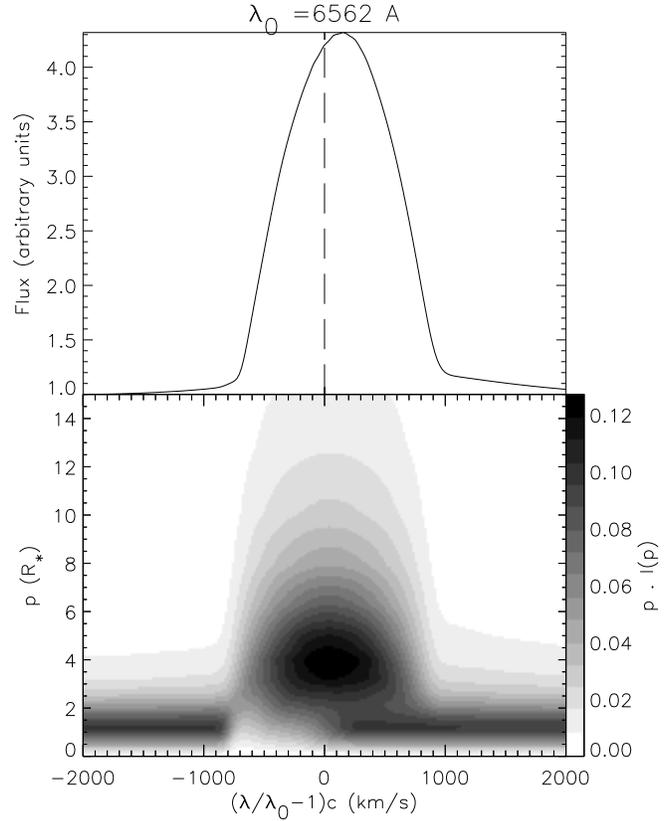, width=8cm}
\caption{{\bf Hot star case (WN8)}
Same set as in Fig.~\ref{fig_jnu2_wn5} but for the H$\alpha$ line in a WN8 model with a terminal
velocity of 840\,km\,s$^{-1}$.
}
\label{fig_jnu2_wn8}
\end{figure}

We are particularly interested in determining where in the outflow
the emergent flux originates from.
Using models for nitrogen-rich Wolf-Rayet (WN) stars similar
to those studied elsewhere (e.g., Hillier \& Millier 1998, Herald
et al. 2001), we show in Figs.~\ref{fig_jnu2_wn5}-\ref{fig_jnu2_wn8}
gray-scale images in the $(x,p)$ plane of the flux-like quantity
$p \cdot I(p)$, where $x = (\lambda/\lambda_0-1)\,c$, $\lambda$ is the
wavelength, $p$ is the impact parameter and $I(p)$ is the specific intensity
(at $\lambda$, or $x$).
Note here that $p$ is in units of the hydrostatic radius $R_{\ast}$ of the WN
star under consideration.
In Fig.~\ref{fig_jnu2_wn5} (WN5 case), we show the case of the resonance line
C{\,\sc iv}\,1548\AA\, in a hot hydrogen-free WN model with an asymptotic
velocity $v_{\infty} = 2000$\,km\,s$^{-1}$, while in Fig.~\ref{fig_jnu2_wn8}
(WN8 case), we show the case of the recombination line H$\alpha$ in a cooler
(hydrogen-rich) WN model (with $v_{\infty} = 840$\,km\,s$^{-1}$).
In each figure, the top panel shows the synthetic line profile, which,
for any scaled velocity $x$, corresponds to the integral over all $p$ of the
quantity $p \cdot I(p)$ plotted in the lower panel.
Thus, for any $x$, we directly identify the contribution at $p$ of the
relative fraction made to the total line profile flux.

The gray-scale illustrations contain a lot of information.
First, beyond the velocity limits of the line (limited naturally to $\pm v_{\infty}$),
we primarily observe continuum photons, although some line photons
are observed outside this range (primarily on the red side) due to
the effects of incoherent electron scattering.
In Fig.~\ref{fig_jnu2_wn5}, the continuum photons originate at $p$ values up to about $p_{\rm lim} = 5
R_{\ast}$, which is the location of the photosphere in this WN star model.
In Fig.~\ref{fig_jnu2_wn8}, we observe a similar effect, although the photosphere is
much closer in, at ca. $2 R_{\ast}$.
But in both cases, the line remains optically-thick out to much larger values of
$p$ -- a few tens of the corresponding $R_{\ast}$.
This is consistent with the generally-accepted fact that line formation in
Wolf-Rayet outflows occurs over large scales compared to the photospheric
radius.


Although the lower panels of Figs.~\ref{fig_jnu2_wn5} and \ref{fig_jnu2_wn8} look
qualitatively similar, Fig.~\ref{fig_jnu2_wn5} shows a P-Cygni profile, while
Fig.~\ref{fig_jnu2_wn8} shows an emission profile.
Close inspection of the lower panels shows that emission beyond $p_{\rm lim}$
is strong, roughly symmetric about line center (modulo turbulence effects):
A key difference is that emission extends up to
$v_{\rm frac} \sim 0.5 v_{\infty}$ for the WN5 model (Fig.~\ref{fig_jnu2_wn5})
but right up to $v_{\rm frac} \sim v_{\infty}$ for the WN8 model
(Fig.~\ref{fig_jnu2_wn8}).
In both cases there is a strong flux deficit on the blue side
for $p < p_{\rm lim}$, which is compensated for in the line profile,
but only up to $\pm$ $v_{\rm frac}$, by the line flux
emitted at larger impact parameters.
For the WN5 case, absorption beyond $-1000$\,km\,s$^{-1}$ is not
compensated by emission and a trough appears in the resulting profile.
For the WN8 case, absorption merely introduces skewness in an otherwise
pure emission profile.

\begin{figure}[htp!]
\epsfig{file=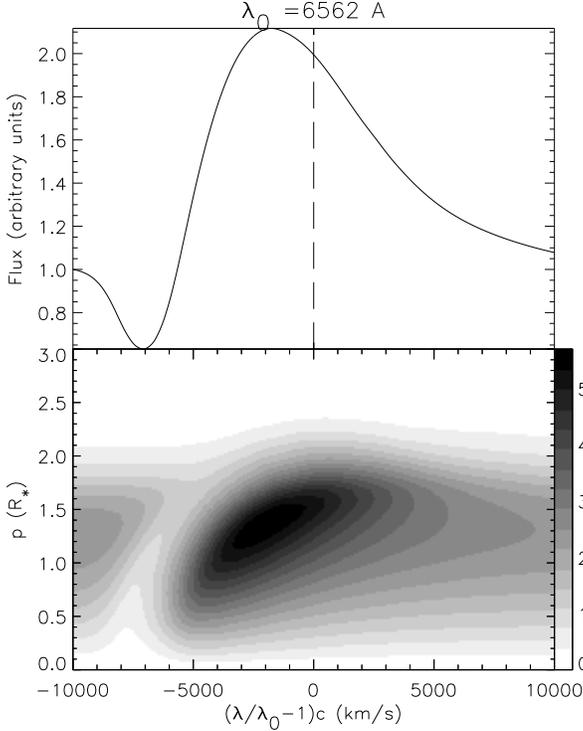, width=8cm}
\caption{{\bf Type II SN model}
Same set as in Figs.~\ref{fig_jnu2_wn5}-\ref{fig_jnu2_wn8}
for the H$\alpha$ line formed in a SN outflow with similar properties
to those described in Sect.~\ref{Sec_int_stage}.
}
\label{fig_jnu2_sn}
\end{figure}

Let us now consider the case of a Type II SN, taking a model
similar to that presented in Sect.~\ref{Sec_int_stage}.
We show the equivalent of Figs.~\ref{fig_jnu2_wn5}-\ref{fig_jnu2_wn8}
for that SN case in Fig.~\ref{fig_jnu2_sn} for H$\alpha$.
$R_{\ast}$ now refers to the base radius of the simulation also called $R_0$.
The difference with the Wolf-Rayet case is large. Line and continua
are confined to similar regions, both becoming optically-thin
beyond $p_{\rm lim}=2 R_{\ast}$.
Interestingly, both emission and absorption occur below $p_{\rm lim}$.
Emission thus only appears on the blue side of the profile, which explains
why the resulting emission profile is strongly blue-shifted.
This occurs because the density drop-off is high in Type II SN
($n=10$ in this model) so that emission is favored in the inner denser
regions, strengthening the impact of disk-occultation.

\begin{figure}[htp!]
\epsfig{file=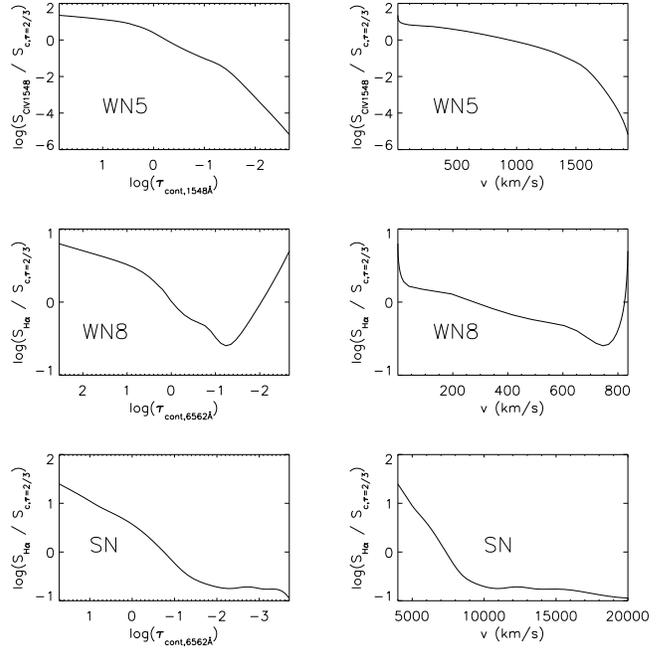, width=9cm}
\caption{Radial and velocity variation of the logarithm of the line source function normalized to the
continuum source function at a continuum optical depth of two third (see text for details).
We show the cases corresponding to Figs.~\ref{fig_jnu2_wn5}-\ref{fig_jnu2_wn8}-\ref{fig_jnu2_sn}, 
i.e. for C{\,\sc iv}\,1548 in the WN5 model (top), H$\alpha$ in the WN8 model (middle) 
and H$\alpha$ in the SN model (bottom).
}
\label{fig_sl_sc}
\end{figure}

What controls the different behaviors shown in
Figs.~\ref{fig_jnu2_wn5}-\ref{fig_jnu2_wn8}-\ref{fig_jnu2_sn}
is the radial variation of the line source function $S_l$ compared to
the continuum source function at a radial optical depth of approximately
$\tau_c=2/3$.  We show in Fig.~\ref{fig_sl_sc} plots of
$S_l$/$S_c(\tau_c=2/3)$ versus
continuum optical depth and versus radial velocity for the three models.
Consider a photospheric (core) ray. The continuum intensity emanating from
that ray is approximately $S_c(\tau_c=1)$ ($\approx S_c(\tau_c=2/3)$), 
where $\tau_c$ now refers to the optical depth
along the ray and we have made use of the Eddington-Barbier relation.
On the other hand, the intensity in an optically thick line at
some frequency in the line will simply be the line source function at
the resonance zone for that frequency. For the WN5 star we see that
the line source function is less than the
reference continuum source function for $\tau_c < 2/3$ and $V > 900$\,\kms,
and drops to very small values for $V>1200$\,\kms.
Thus we expect no emission (as compared to the continuum) for $V >
1200\,$\kms, as indeed seen in Fig.~\ref{fig_jnu2_wn5}.
For the WN8 star the H$\alpha$ source function is less than
$S_c(\tau_c=2/3)$ beyond 300\,\kms, but it drops, by at most a factor of 4
at 750\,\kms.
Thus the flux is reduced relative to the continuum, but never goes black.
The reduced flux is easily compensated by emission from rays not striking
the photosphere, and hence we see no P-Cygni profile.

The SN model is different yet again. From Fig.~\ref{fig_sl_sc}, we see
the line source function
drops below the continuum source function for velocities greater than
7000\,\kms.
Thus, for such velocities, we expect to see reduced emission, as seen from
Fig.~\ref{fig_jnu2_sn}. Further, because the continuum opacity is
mainly due to electron scattering, we can see emission arising from
optical depths greater than $\tau_c=1$ ($V < 7000$\,\kms). Because the line
source
function is larger than the continuum in this region, the line is in emission.
This is again seen in Fig.~\ref{fig_jnu2_sn}. Unlike the two WR models,
we get significant emission for rays striking the photosphere.
Because (by definition) the photosphere is thick we do not see
much emission from the backside of the flow. Consequently the emission,
for photospheric rays, is strongly skewed to the blue (Fig.~\ref{fig_jnu2_sn}).

\begin{figure}[htp!]
\epsfig{file=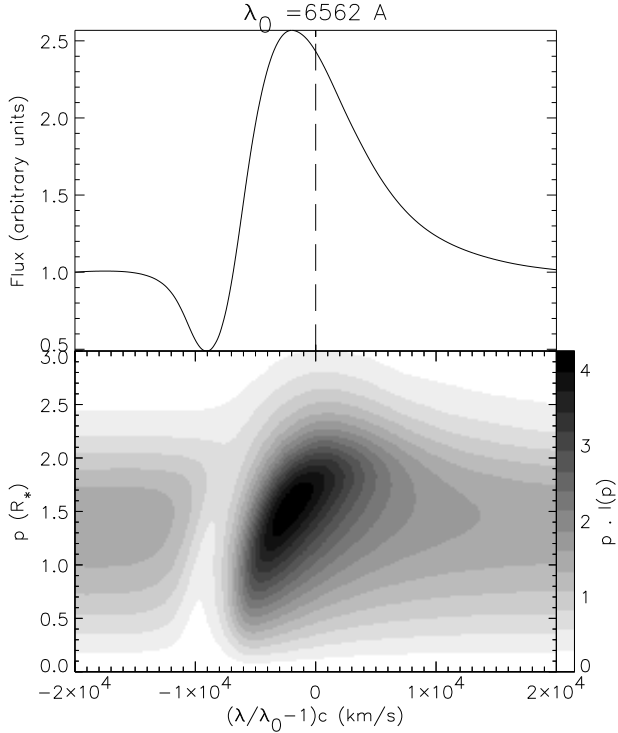 , width=8cm}
\caption{
Same as Fig.~\ref{fig_jnu2_sn} for a model with $n=8$.
}
\label{fig_jnu2_sn_n8}
\end{figure}

\begin{figure}[htp!]
\epsfig{file=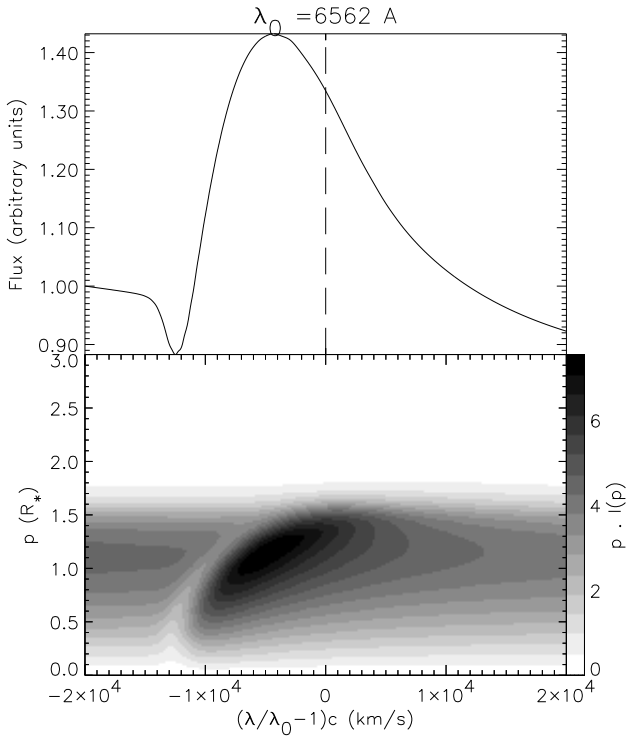, width=8cm}
\caption{Same as Fig.\ref{fig_jnu2_sn} but for a model with $n=14$.
}
\label{fig_jnu2_sn_n14}
\end{figure}

Finally, we show in Figs.~\ref{fig_jnu2_sn_n8} and \ref{fig_jnu2_sn_n14}
the cases for SN models that have different density
exponents, i.e. $n=8$ and $n=14$ respectively compared to the model shown
in Fig.~\ref{fig_jnu2_sn}.
Indeed, the higher $n$ is, the faster the density drops and the more
blue-shifted the resulting H$\alpha$ profile looks.
While the source function follows a similar trend in these two additional
cases (not shown), it is the modulated influence of disk occultation that gives
rise to different blue-shift magnitudes in the resulting profile.
Overall, the large density exponent characterizing the early evolution
of Type II SN is also a key ingredient that precludes the possibility of
observing a clear departure from sphericity (if at all present), the
line and continuum formation regions overlapping in space and being so confined.

\section{Conclusion}
\label{Sec_conc}

We have presented first results based on the model atmosphere code CMFGEN
(Hillier \& Miller 1998) for the spectroscopic modeling of photospheric-phase
Type II supernovae. To facilitate the modeling, several changes were made to
the code: these include the ability to use
a Hubble velocity law and a power-law density distribution,
an adaptive grid to handle the steep H recombination front in cooler
SN models, and an improved algorithm to compute the gray temperature
distribution in a rapidly expanding outflow.
Still to be included in our treatment are relativistic effects, energy deposition
from radioactive decay, and the possibility to base our computation on a
computed hydrodynamical structure.
We demonstrated the ability of CMFGEN to reproduce a large number of features
in high S/N multi-wavelength spectroscopic observations of Type II SN, using
the very well-observed SN1987A and SN1999em.
The observed spectral energy distribution from UV to the near-IR (when available)
can be matched very accurately: the fit quality is in general within the 10\% level,
although it is difficult to match observations in the UV range to better than 20-30\%.
We show examples taken over a wide range of outflow ionization properties, covering
fully- (early-time) and partially- (late-time) ionized spectra.
This good behavior of CMFGEN gives us hope that a wealth of information can
be retrieved through a detailed analysis of multi-epoch observations of Type II SN.
Based on a representative set of observations, we show that He{\,\sc i} lines in early-time
spectra can be well fitted by invoking modest helium enrichment typical of an
OB-supergiant progenitor.
The non-LTE radiative transfer solution also predicts the presence of lines in the optical
from higher ionization than before expected, in particular Fe{\,\sc iii} and N{\,\sc ii}.
The latter provides a simple explanation for the appearance of ``bumps'' in the
blue-wing of H$\beta$ and He{\,\sc i}\,5875\AA, generally interpreted in terms of over-dense emitting
(in the corresponding line) structures lying far above the photosphere.
Overall, this study shows that taking into account as accurately as possible the enormous
effects of line-blanketing is key for a detailed modeling of Type II SN.
We also discuss through a number of different stand-points the problem of line and
continuum formation in Type II SN, giving vivid illustrations for both.
In particular, for the first time, we give an illustration to explain the
sometimes strong blue-ward wavelength shift of, e.g., H$\alpha$ in Type II SN.
The shift arises from the combined effects of disk-occultation and a line formation
region confined to the photosphere.

Forthcoming studies will present more focused analyses on specific issues.
Since most investigations done on Type II SN are with the aim of using them for
the distance calibration of the Universe (as holds for Ia's),
we will next study the Expanding Photosphere Method
and the analysis of Eastman et al. (1996), following their approach but using CMFGEN
to search for possible differences, in, e.g., correction factors.
We will also discuss methods for determining the photospheric velocity,
an issue of importance since the rate of expansion of the radiating
photosphere is one of the key quantities entering any such distance
determination.
Long term goals are to design a very reliable method to determine distances in the
Universe based on Type II SN, with application to a sample of very well observed objects.
Analyses of individual objects is also sought, in order to constrain the
evolutionary status of the SN progenitor based on a {\it spectroscopic} approach,
to complement inferences made exclusively from photometric measurements performed on the
pre-explosion object (see, e.g., Smarrt et al. 2004).


\end{document}